\documentclass{article}
\usepackage[titletoc,title]{appendix}
\usepackage{graphicx}
\usepackage{epstopdf}
\usepackage{amsmath, amsthm, latexsym}
\usepackage{amssymb}
\usepackage{slashed}

\DeclareMathOperator{\sech}{sech}
\DeclareMathOperator{\tr}{tr}

\usepackage[small, bf]{caption}
\usepackage{subfigure}
\usepackage{hyperref}

\newtheorem{theorem}{Theorem}

\newtheorem{remark}{Remark}

\usepackage[usenames,dvipsnames]{xcolor}
\definecolor{candyapplered}{rgb}{1.0, 0.03, 0.0}
\hypersetup{colorlinks, linkcolor={red!70!black}, citecolor={cyan}, urlcolor={blue!80!black}}

\numberwithin{equation}{section}

      {\arraycolsep 0.14em\begin{eqnarray}}{\end{eqnarray}}

\begin{document}

\begin{titlepage}\begin{center}
\textbf{\Large{Dirac particles in a gravitational shock wave}}\\\vspace{6mm}

\textbf{\normalsize Peter Collas}\footnote{Department of Physics and Astronomy, California State University, Northridge, Northridge, CA 91330-8268. Email: peter.collas@csun.edu.}
\textbf{\normalsize and David Klein}\footnote{Department of Mathematics and Interdisciplinary Research Institute for the Sciences, California State University, Northridge, Northridge, CA 91330-8313. Email: david.klein@csun.edu.}\\
\end{center}

\begin{center} (April 2018) \end{center}

\begin{center} \href{https://doi.org/10.1088/1361-6382/aac144}{Class. Quantum Grav. \textbf{35}, 125006 (2018)} \end{center}
\vspace{2cm}

\begin{abstract}

\noindent Using the Newman-Penrose formalism we calculate the positive energy momentum eigenstates of the Dirac equation for a plane polarized gravitational wave pulse. We then consider Dirac particles whose spins are polarized in each orthonormal coordinate direction prior to the arrival of the gravitational wave pulse, and determine how the spins are changed by the pulse after its departure.
\end{abstract}
\vspace{1cm}

\end{titlepage}

\newpage

\section[Introduction]{Introduction}\label{intro}

\noindent

\noindent Ever since Chandrasekhar's solution in Kerr spacetime \cite{C76}, the Dirac equation has been used extensively to study the influence of curved spacetimes on the quantum mechanical behavior of particles, for example \cite{Carter, P80,S91,Z96,FR09,HP09,R17}. An excellent reference on quantum field theory in curved spacetime is Parker and Toms text \cite{PT09}.\\

\noindent  Detailed analyses of spinning objects in general relativity for the classical case (non quantum) were carried out by Dixon \cite{D70I,D70II,D74III}.  Later Hehl and Ni made use of the Foldy-Wouthyusen representation to examine the behavior of a Dirac particle in a non-inertial frame in Minkowski spacetime \cite{HN90}.  Since then several authors have used the Foldy-Wouthyusen approach to relate the quantum behavior of Dirac particles to the behavior of classical spinning particles in various general relativistic backgrounds, for example \cite{S08,MO13,OST13} and references therein.\\  

\noindent Of particular interest is the effect of gravitational waves on quantum particles with spin.  Classical spinning particles in gravitational wave backgrounds, in the weak-field approximation, were considered in \cite{MS00,M02} and recently in \cite{BGO17}, but to our knowledge there has not been an investigation of a Dirac particle in an \emph{exact} gravitational wave.  This may be due in part to the non-existence of a conserved spin operator in curved spacetime \cite{R01}.\\ 

\begin{figure}[!h]
  \begin{center}
    \includegraphics[width=2.5in]{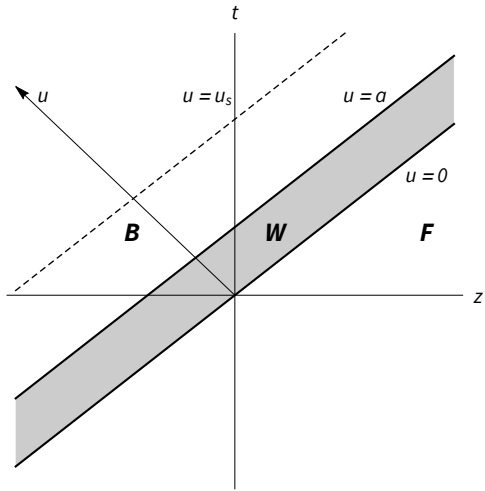}
     \end{center}
   \caption{The `sandwich' wave geometry. The shaded strip, $W$ represents the gravitational wave pulse. To the right of $W$ is region $F$, the front of the wave consisting of spacetime points not yet affected by the wave pulse, and to the left is region $B$, the back of the wave.  The $t$-axis is the Fermi observer timelike geodesic.}
  \label{GSWFig1}
\end{figure}

\noindent In the present paper we overcome this difficulty for a particular gravitational wave spacetime.  We consider the spacetime of a gravitational pulse or ``sandwich'' wave. For background on spacetimes of this type, we refer the reader to \cite{BPR59, R86,BFI89, BP89, R06,GP09}. The spacetime $(\mathcal{M}, \mathrm{g})$ that we consider is a union of three regions, as shown in Figure \ref{GSWFig1}.  The first region, or front part, $F$, is flat. The wave pulse, which travels in the positive $z$-direction, has not yet reached the spacetime points within it. The next region is the pulse or wave part, $W$.  It has non vanishing Weyl tensor but is Ricci flat.  The remaining region, or back part, $B$, is again flat and consists of all spacetime points that the wave pulse has already passed.\\

\noindent A single coordinate chart $\{t,x,y,z\}$ covers the spacetime $\mathcal{M}$ (and hence all three regions), and the metric tensor $\mathrm{g}$ expressed in those coordinates has the form given by the line element of Eq.\eqref{m1}.  The metric $\mathrm{g}$  satisfies the vacuum field equations (see Eq. \eqref{m4.1}) within all three regions, $F, W, B$, and it is of class $C^{1}$ on $\mathcal{M}$ (and is smooth within each of the three regions)\footnote{There is a quasiregular singularity in the region $B$ of  $\mathcal{M}$, but this does not affect our analysis; see Remark \ref{remarksing}.}\\   

\noindent We use the Newman-Penrose formalism to solve the Dirac equation on $\mathcal{M}$ for the positive energy momentum eigenstates and find explicit formulas for their restrictions to each of the three regions.  At this stage of our analysis, the tetrad field for the spinor solutions to the Dirac equation is determined in a natural way by the global coordinate chart, $\{t,x,y,z\}$.  In region $F$ (in front of the wave), the metric expressed in  $\{t,x,y,z\}$ coordinates is Minkowskian and the tetrad field is the canonical one for Minkowski spacetime.   Since there exists a well-defined spin operator in Minkowski spacetime, we can establish an initial spin polarization to the spinor solution of the Dirac equation restricted to region $F$, and we do so in the reference frame of the Dirac particle (i.e., with zero $3$-momentum).\\ 

\noindent In order to analyze the effect of the shock wave on the initially polarized spinors, it is necessary to compare the spin of the Dirac particle in region $B$, to the original polarization in region $F$ in a coherent way, but here a difficulty arises.   Although the post wave region $B$ is flat, the metric $\mathrm{g}$ expressed in the global $\{t,x,y,z\}$ coordinates is not Minkowskian, and there are infinitely many coordinate transformations that express $\mathrm{g}$ in Minkowskian form, each with an associated canonical tetrad field in region $B$ (they are all related by Lorentz transformations).  Regions $B$ and $F$ are the disjoint connected components of the flat portion of $\mathcal{M}$, and so the Minkowski coordinate patch in region $F$ cannot be uniquely extended to $B$.\\

\noindent To find the appropriate Minkowski coordinate system and associated tetrad field in region $B$, so that pre- and post-spin calculations can be meaningfully compared, we first find Fermi coordinates along a timelike geodesic $\gamma(\tau)$ (where $\tau$ is proper time) that begins in region $F$, goes through the wave pulse $W$, and ends in region $B$.  The path $\gamma(\tau)$, which we describe as the ``laboratory frame,'' is chosen to remain at a fixed spatial location with respect to the Minkowski chart in $F$ (see Figure \ref{GSWFig1}).  The Fermi coordinates are then uniquely determined by the parallel transport of the Minkowski tetrad in $F$ along $\gamma(\tau)$.  \\

\noindent The Fermi coordinates defined in this way are Minkowskian in both regions $F$ and $B$.  We derive the diffeomorphism from the initial (global) coordinates $\{t,x,y,z\}$ restricted to $B$, to the new Fermi coordinates $\{T,X,Y,Z\}$ in $B$  (which are Minkowskian).  We then change to the new Minkowski coordinates in $B$ and carry out a tetrad rotation from the original tetrad field $h_A{}^\alpha{}$ to the canonical Minkowski tetrad field $\hat{e}_A{}^\alpha{}$ using a local Lorentz transformation $\Lambda$ in $B$.\\ 

\noindent The local Lorentz transformation $\Lambda$ determines a transformation matrix $L$ according to which our spinor solution to the Dirac equation is transformed when the tetrad field is tranformed by $\Lambda$.  We then have the relationship $\Psi (\hat{e})=L\cdot\psi_{BFermi}(h)$, where $\Psi(\hat{e})$, is the new transformed spinor in region $B$ in exact Fermi coordinates and with the canonical Minkowski (Fermi) tetrad $\hat{e}$, and $\psi_{BFermi}(h)$ is the original Dirac spinor restricted to region $B$ relative to the original tetrad field $h$ associated with the coordinates $\{t,x,y,z\}$, but whose components are functions of $\{T,X,Y,Z\}$.  A calculation then shows that  $\Psi$ is not an eigenvector of the Minkowski spin operator in region $B$.  It then follows, from a theorem we introduce and prove (Theorem \ref{theorem}), that our spinor solution to the Dirac equation in any possible Minkowski tetrad field is not an eigenvector to the stationary spin operator for that tetrad field in region $B$. Therefore the Dirac particle cannot be stationary and polarized in any of the infinitely many possible Minkowski charts in region $B$.  The gravity wave pulse in this way disturbs the spins of the Dirac particle.  \\ 

\noindent Our paper is organized as follows.  In Sec. \ref{N-P} we introduce the metric and derive the Dirac equations for the spinor components using the Newman-Penrose formalism.  In Sec. \ref{S} we separate the variables and find the general solutions.  In Sec. \ref{sand} we provide some background information about the behavior of classical dust particles in the specific spacetime that we study.  We then find the solutions for each of the three regions.  Since our metric is of class $C^{1}$ it follows that the resulting solutions are also $C^{1}$.  In Sec. \ref{sing} we parallel transport the Minkowski tetrad from region $F$ to region $B$ along the Fermi observer's worldline, and remove the coordinate singularity in region $B$ by finding exact Fermi coordinates and transformation formulas.  We explicitly show that these coordinates are Minkowskian.  In Sec. \ref{sing2}, we give the local Lorentz transformation $\Lambda$ that rotates of the tetrad field into a canonical Minkowski tetrad field, and transform the spinor solution to the Dirac equation accordingly.  In Sec. \ref{spin} we state and prove our Theorem \ref{theorem} on Lorentz transformations of the spin operators and compare the polarized spinors of the solution in region $F$ to the spinors in region $B$ and show that the gravitational wave pulse depolarizes the Dirac particle in any Minkowski tetrad field.  Sec. \ref{conc} summarizes and discusses our results.  The four Appendices  follow the main exposition and provide details of the various calculations.

\section[The Dirac equation in the N-P formalism]{The Dirac equation in the N-P formalism} \label{N-P}

\noindent In all that follows we adopt Planck units so that, $c=G=\hbar=1$. We consider a transverse plane polarized gravitational wave propagating in the z direction.  The line element for such a wave may be written as shown below \cite{R86}, \cite{R06}, \cite{GP09},
\begin{equation}
\label{m1}
ds^{2}=dt^{2}-dz^{2}-f^{2}(u)dx^{2}-g^{2}(u)dy^{2}\,,
\end{equation}

\noindent where
\begin{equation}
\label{m2}
u=\frac{1}{\sqrt{2}}(t-z)\,.
\end{equation}

\noindent Introducing the variable
\begin{equation}
\label{m3}
v=\frac{1}{\sqrt{2}}(t+z)\,,
\end{equation}

\noindent we transform the metric to
\begin{equation}
\label{m4}
ds^{2}=2dudv-f^{2}(u)dx^{2}-g^{2}(u)dy^{2}\,.
\end{equation}

\noindent The functions $f(u)$ and $g(u)$ must satisfy the only nontrivial vacuum field equation, namely,
\begin{equation}
\label{m4.1}
R_{uu}=-\left(\frac{f^{\prime\prime}(u)}{f(u)}+\frac{g^{\prime\prime}(u)}{g(u)}\right)=0\,.
\end{equation}

\begin{remark}\label{bptock}
One may use a coordinate transformation to recast the Bondi-Pirani plane wave, \textup{Eq.(2.1), of ref. \cite{BP89}}, into the form of \textup{Eq. (17.11)} of Griffiths-Podolsk\'{y} \textup{\cite{GP09}} \textup{(}with $\omega=0$\textup{)} and hence to the form \textup{Eq.} \eqref{m4}.
\end{remark}

\noindent  We use the Newman-Penrose (N-P) formalism to write and solve the Dirac equation in the spacetime of Eq. \eqref{m4}.  A set of orthogonal tetrad vectors for the metric of Eq. \eqref{m4} is given below for which we adopt the ordering convention $e_A{}=(e_A{}^x{},e_A{}^y{},e_A{}^v{},e_A{}^u{})$.
\begin{align}
e_1{}&=\left(\frac{1}{f(u)},\,0,\,0,\,0\right)=\frac{1}{f(u)} \partial_x\,,\label{m7}\\\nonumber\\
e_2{}&=\left(0,\,\frac{1}{g(u)},\,0,\,0\right)=\frac{1}{g(u)} \partial_y\,,\label{m8}\\\nonumber\\
e_3{}&=\left(0,\,0,\,\frac{1}{\sqrt{2}},\,-\frac{1}{\sqrt{2}}\right)=\frac{1}{\sqrt{2}}\partial_v-\frac{1}{\sqrt{2}} \partial_u\,,\label{m9}\\\nonumber\\
e_4{}&=\left(0,\,0,\,\frac{1}{\sqrt{2}},\,\frac{1}{\sqrt{2}}\right)=\frac{1}{\sqrt{2}}\partial_v+\frac{1}{\sqrt{2}} \partial_u\,.\label{m10}
\end{align}

\noindent From these vectors we construct the complex null tetrad $\{l, n, m, \overline{m}\}$ as follows \cite{GHP73}:
\begin{align}  
\lambda_{1}&=l=\frac{1}{\sqrt{2}}\left(e_4{}+e_3{}\right)=\frac{1}{\sqrt{2}}\left(0,\,0,\,\sqrt{2},\,0\right)\,,\label{m11}\\\nonumber\\
\lambda_{2}&=n=\frac{1}{\sqrt{2}}\left(e_4{}-e_3{}\right)=\frac{1}{\sqrt{2}}\left(0,\,0,\,0,\,\sqrt{2}\right)\,,\label{m12}\\\nonumber\\
\lambda_{3}&=m=\frac{1}{\sqrt{2}}\left(e_1{}+ie_2{}\right)=\frac{1}{\sqrt{2}}\left(\frac{1}{f(u)},\,i\frac{1}{g(u)},\,0,\,0\right)\,,\label{m13}\\\nonumber\\
\lambda_{4}&=\overline{m}=\frac{1}{\sqrt{2}}\left(e_1{}-ie_2{}\right)=\frac{1}{\sqrt{2}}\left(\frac{1}{f(u)},\,-i\frac{1}{g(u)},\,0,\,0\right).\label{m14}
\end{align}

\noindent We write the Dirac equation in the N-P formalism using the standard notation for the spin coefficients, \cite{C76}, \cite{C92} as follows:
\begin{align}
(\Delta+\mu^{*}-\gamma^{*})G_{1}-(\delta^{*}+\beta^{*}-\tau^{*})G_{2}&=i\mu_{*}F_{1}\,,\nonumber\\\nonumber\\
(D+\varepsilon^{*}-\rho^{*})G_{2}-(\delta+\pi^{*}-\alpha^{*})G_{1}&=i\mu_{*}F_{2}\,,\nonumber\\\label{d1}\\
(D+\varepsilon-\rho)F_{1}+(\delta^{*}+\pi-\alpha)F_{2}&=i\mu_{*}G_{1}\,,\nonumber\\\nonumber\\
(\Delta+\mu-\gamma)F_{2}+(\delta+\beta-\tau)F_{1}&=i\mu_{*}G_{2}\,,\nonumber
\end{align}

\noindent where $\mu_{*}\sqrt{2}=m=\mbox{the mass of the particle}$ (not to be confused with the null vector $m$ in Eq. \eqref{m13}), and the spinor $\psi(x,y,v,u)$ has the form:
\begin{equation}
\label{d1.1}
\psi=\left(\begin{array}{c}F_1{}\\
F_2{}\\
G_1{}\\
G_2{}
\end{array}\right).
\end{equation}

\noindent We use the standard notation to designate the null tetrad vectors as directional derivatives
\begin{equation}
\label{d2}
\lambda_1{}=l=D,\;\;\lambda_2{}=n=\Delta,\;\;\lambda_3{}=m=\delta,\;\;\lambda_4{}=\overline{m}=\delta^{*}\,.
\end{equation}

\noindent From these we have the four scalar operators,
\begin{equation}
\label{d3}
D=l^\alpha{}\nabla_\alpha{},\;\;\Delta =n^\alpha{}\nabla_{\alpha},\;\;\delta=m^\alpha{}\nabla_{\alpha},\;\;\delta^{*}=\overline{m}^\alpha{}\nabla_{\alpha}\,,
\end{equation}

\noindent and for any scalar function $h$ we write
\begin{equation}
\label{d4}
Dh=l^\alpha{}h_{,\alpha},\;\;\Delta h=n^\alpha{}h_{,\alpha},\;\;\delta h=m^\alpha{}h_{,\alpha},\;\;\delta^{*}h=\overline{m}^\alpha{}h_{,\alpha}\,.
\end{equation}

\noindent In our case the Dirac equation Eq. \eqref{d1} reduces to,
\begin{align}
(\Delta+\mu)G_{1}-\delta^{*}G_{2}&=i\mu_{*}F_{1}\,,\nonumber\\\nonumber\\
DG_{2}-\delta G_{1}&=i\mu_{*}F_{2}\,,\nonumber\\\label{d6}\\
DF_{1}+\delta^{*}F_{2}&=i\mu_{*}G_{1}\,,\nonumber\\\nonumber\\
(\Delta+\mu)F_{2}+\delta F_{1}&=i\mu_{*}G_{2}\,.\nonumber
\end{align}

\section[Separation of variables and the solutions]{Separation of variables and the solutions} \label{S}

\noindent In working out the details of Eqs. \eqref{d6}, first we cancel out all the $1/ \sqrt{2}$ factors appearing on both sides of the equations (recall that $\mu_{*}\sqrt{2}=m$).  Since the metric in Eq. \eqref{m4} does not depend on $x, y,$ and $v$, the corresponding canonical momenta are constants, both in classical and quantum mechanics.  We take advantage of this fact and write the 4-component spinor $\psi$ as
\begin{equation}
\label{s1}
\psi(x,y,v,u)=\left(\begin{array}{c}F_1{}\\
F_2{}\\
G_1{}\\
G_2{}
\end{array}\right)=e^{-i(p_x{}x+p_y{}y+p_v{}v)}
\left(\begin{array}{c}\mathbb{F}_1{}(u)\\
\mathbb{F}_2{}(u)\\
\mathbb{G}_1{}(u)\\
\mathbb{G}_2{}(u)
\end{array}\right),
\end{equation}

\noindent where, $p_A{},\;A=x,y,v$, are the eigenvalues of the operators $i\partial_A{}$.  Thus $\psi$ is an eigenfunction of the canonical momenta $p_A{}$ (see Sec. \ref{F}).

\begin{remark}\label{conp}
If we write the metric in the coordinates of Eq. \eqref{m1}, then the conserved $p_v{}$ is given by
\begin{equation}
\label{s1a}
p_v{}=\frac{p_t{}+p_z{}}{\sqrt{2}}\,,
\end{equation}

\noindent where the non-conserved $p_t{}$ and $p_z{}$, can be obtained from the Lagrangian corresponding to the metric \eqref{m1}.
\end{remark}

\noindent The resulting equations for the $\mathbb{F}_{i}(u)$ and the $\mathbb{G}_{i}(u)$ are
\begin{align}
\left(\frac{ip_x{}}{f(u)}+\frac{p_y{}}{g(u)}\right)\mathbb{G}_{2}+\sqrt{2}\,\mu\,\mathbb{G}_{1}+\sqrt{2}\,\mathbb{G}^{\,\prime}_{1}&=im\mathbb{F}_{1}\,,\label{s2}\\\nonumber\\
\left(\frac{ip_x{}}{f(u)}-\frac{p_y{}}{g(u)}\right)\mathbb{G}_{1}-i\sqrt{2}\,p_v{}\,\mathbb{G}_{2}&=im\mathbb{F}_{2}\,,\label{s3}\\\nonumber\\
-\left(\frac{ip_x{}}{f(u)}+\frac{p_y{}}{g(u)}\right)\mathbb{F}_{2}-i\sqrt{2}\,p_v{}\,\mathbb{F}_{1}&=im\mathbb{G}_{1}\,,\label{s4}\\\nonumber\\
\left(-\frac{ip_x{}}{f(u)}+\frac{p_y{}}{g(u)}\right)\mathbb{F}_{1}+\sqrt{2}\,\mu\,\mathbb{F}_{2}+\sqrt{2}\,\mathbb{F}^{\,\prime}_{2}&=im\mathbb{G}_{2}\,,\label{s5}
\end{align}

\noindent where the spin coefficient $\mu$ is given by
\begin{equation}
\label{s5a}
\mu=\frac{1}{2}\left(\frac{f^{\prime}(u)}{f(u)}+\frac{g^{\prime}(u)}{g(u)}\right)\,.
\end{equation}

\noindent We now solve for $\mathbb{F}_{1}(u)$ and $\mathbb{G}_{2}(u)$ using the algebraic Eqs. \eqref{s3} and \eqref{s4}, and substitute the results in the differential Eqs. \eqref{s2} and \eqref{s5} obtaining two decoupled and \textit{identical} differential equations for $\mathbb{G}_{1}(u)$ and $\mathbb{F}_{2}(u)$,
\begin{align}
\left[i\left(m^{2}f^{2}g^{2}+p_{x}^{2}\,g^{2}+p_{y}^{2}\,f^{2}\right)+p_v{}fg\left(gf^{\prime}+fg^{\prime}\right)\right]\mathbb{G}_{1}(u)+\nonumber\\2p_v{}\,f^{2}g^{2}\,\mathbb{G}^{\,\prime}_{1}(u)&=0\,,\label{s6}\\\nonumber\\
\left[i\left(m^{2}f^{2}g^{2}+p_{x}^{2}\,g^{2}+p_{y}^{2}\,f^{2}\right)+p_v{}fg\left(gf^{\prime}+fg^{\prime}\right)\right]\mathbb{F}_{2}(u)+\nonumber\\2p_v{}\,f^{2}g^{2}\,\mathbb{F}^{\,\prime}_{2}(u)&=0\,.\label{s7}
\end{align}

\noindent Given explicit expressions for the functions $f(u)$ and $g(u)$, the general solutions to the Eqs. \eqref{s6} and \eqref{s7}, can be obtained by performing the integration
\begin{align}
\label{s8}
\frac{\mathbb{G}_{1}(u)}{b_{1}}&=\frac{\mathbb{F}_{2}(u)}{a_{2}}=\nonumber\\
&\exp{\left[\int_{u_{0}}^{u}\frac{-i\left(p_x{}^{2}g^{2}+p_y{}^{2}f^{2}+m^{2}f^{2}g^{2}\right)-p_v{}fg\left(fg^{\prime}+gf^{\prime}\right)}{2p_v{}f^{2}g^{2}}dw\right]}\,,
\end{align}

\noindent where in the integrand $f=f(w)$, $g=g(w)$.  The constants $a_{2}$ and $b_{1}$, are chosen in region $F$ and they determine the polarization of the particle in its rest frame as we show in Sec. \ref{spin}.  They depend on $m,\,p_x{},\,p_y{},\, p_v{},\,p_u{}$ and as we will see below, they must remain the same through all the three regions of our shock wave spacetime, in order for the different parts of the wavefunction to match properly at the shock wave boundaries, i.e., for continuous differentiability of the spinor $\psi$ to hold.

\section[The `sandwich' wave]{The sandwich wave} \label{sand}

\noindent In general one may choose an arbitrary function $g(u)$ and solve Eq. \eqref{m4.1} for $f(u)$.  In this section, however, we shall solve Eq. \eqref{s8} for a \textit{sandwich} gravitational wave \cite{R86, BFI89, BP89, R06,GP09}. As described in the introduction and illustrated in Figure \ref{GSWFig1}, the spacetime $\mathcal{M}$ is a union of three regions, i.e.,
\begin{equation}
\mathcal{M}=F\cup W \cup B,
\end{equation}
where $B$ represents the \textit{back} or post-wave region, $W$ represents the \textit{wave} pulse, and $F$ represents the \textit{front} of the wave, that is, the points in advance of the wave.  Each of these regions is characterized by the variable $u= (t-z)/\sqrt{2}$, as indicated below.  The functions $f(u)$ and $g(u)$ which appear in the metric Eq. \eqref{m4} are now defined by,
\begin{equation}
\label{sw1}
f(u)=
\begin{cases}
1, &u\in(-\infty,0]\equiv F,\\
\cos{(ku)}, &u\in[0,a]\equiv W,\\
\cos{(ka)}+k(a-u)\sin{(ka)}, &u\in[a,\infty)\equiv B,\\
\end{cases}
\end{equation}
\begin{equation}
\label{sw2}
\qquad g(u)=
\begin{cases}
1, &u\in(-\infty,0]\equiv F,\\
\cosh{(ku)}, &u\in[0,a]\equiv W,\\
\cosh{(ka)}+k(u-a)\sinh{(ka)}, &u\in[a,\infty)\equiv B,\\
\end{cases}
\end{equation}\\

\noindent where $a>0$ and $k$ are constants.  The functions $f(u)$ and $g(u)$ satisfy the vacuum Eq. \eqref{m4.1} in all three regions.  The resulting composite metric $\mathrm{g}$ is of class $C^{1}$. Using Eq. \eqref{m2} it is easy to see that the plane wave `slab,' $W$, consisting of the spacetime points satisfying $u\in[0,a]$, is moving in the positive $z$-direction as $t$ increases.  Since $f$ and $g$ are linear functions of $u$ on $F$ and $B$, we have $f^{\prime\prime}(u)=g^{\prime\prime}(u)=0$, so all the curvature components, $R_{\alpha\beta\gamma\delta}$, vanish in these regions.  In light of this we define the limit $k\rightarrow0$, as the flat spacetime limit, wherein the shock wave disappears.\\  

\noindent We shall choose $a$ and $k$ so that $ka\in(-\pi/2,\pi/2)$ thus $f(u)>0$ in region $W$.  However, $f(u)$ necessarily vanishes at one point, $u=u_{s}$, in region $B$. We deal with this singularity in Section \ref{sing}.\\

\noindent Before turning to the solution to the Dirac equation for this spacetime, we briefly describe the effect of the wave pulse on classical test particles in analogy to the classical ``sticky bead'' problem. Similar descriptions for spacetimes of this type are given in \cite{R86,R06} and systematic analyses of timelike geodesics in wave pulse spacetimes were carried out by Penrose and other researchers, cf., \cite{GP09}.\\

\begin{figure}[!h]
  \begin{center}
    \includegraphics[width=2.5in]{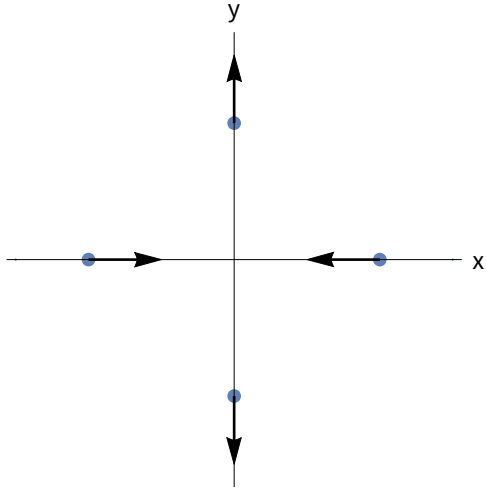}
     \end{center}
   \caption{Motion of classical test particles in the wave pulse.}
  \label{GSWFig2}
\end{figure}

\noindent It is easily verified that the equations, $x,y,z= \text{constant}$, describe timelike geodesics with affine parameter $t$.  Consider two dust particles at rest in region $F$ before the wave pulse arrives, separated only by a small $x$-difference $dx$.  The proper distance between them for all times is $ds=f(u)dx$, where $dx$ is constant.  Similarly two dust particles separated only by a small $y$-distance $dy$ have proper distance between them $ds=g(u)dy$.  Thus, as depicted in Figure \ref{GSWFig2}, during the passage of the wave, the $x$-separation of the particles decreases and the $y$-separation increases since $f(u)=\cos u$ decreases and $g(u)=\cosh u$ increases, as $u$ increases from zero.

\noindent Returning to the Dirac equation, for convenience, we write the argument of the exponential in Eq. \eqref{s8} as
\begin{equation}
\label{chi2}
\Phi(u_{0},u)=\int_{u_{0}}^{u}\phi(w)dw\,,
\end{equation}

 \noindent where the lower limit of integration, $u_{0}$, is chosen so that no singularity of the integrand $\phi(w)$ lies between $u_{0}$ and $u$, but is otherwise arbitrary.  In what follows it is convenient to choose $u_{0}=0$.  Since the integrand of Eq. \eqref{s8} contains terms of class $C^{0}$ the spinors will be of class $C^{1}$.

\subsection[The solution in region $F$]{The solution in region $F$}\label{F}

\noindent We begin with the solution in region $F$.  Evaluating Eq. \eqref{s8}, with the appropriate choices from the definitions \eqref{sw1}, \eqref{sw2}, we obtain,
\begin{equation}
\label{B1}
\Phi_{F}(0,u)=-i\frac{\left(m^{2}+p_x{}^{2}+p_y{}^{2}\right)}{2p_v{}}\,u=-ip_u{}u.
\end{equation}

\noindent Since in region $F$, $p_{u}$ and $p_t{}$ are conserved, we have the useful relations below,
\begin{equation}
\label{B2}
m=\sqrt{2p_v{}p_u{}-p_{x}^{2}-p_{y}^{2}}=\sqrt{p_{t}^{2}-\left(p_{x}^{2}+p_{y}^{2}+p_{z}^{2}\right)},
\end{equation}

\begin{align}
p_\mu{}x^\mu{}&=p_x{}x+p_y{}y+p_v{}v+p_u{}u,\nonumber\\
&=p_x{}x+p_y{}y+p_z{}z+p_t{}t,\label{B3}
\end{align}

\noindent and
\begin{align}
p_t{}+p_z{}&=\sqrt{2}\;p_v{},\label{B3a}\\
p_t{}-p_z{}&=\sqrt{2}\;p_u{}.\label{B3b}
\end{align}

\noindent Thus
\begin{align}
\mathbb{F}_{2F}(u)&=a_{2}\,e^{\Phi_{F}(0,u)}\,,\label{B4}\\\nonumber\\
\mathbb{G}_{1F}(u)&=b_{1}\,e^{\Phi_{F}(0,u)}\,.\label{B5}
\end{align}

\noindent Then Eqs. \eqref{s3} and \eqref{s4} give us the $\mathbb{F}_{1F}(u)$ and $\mathbb{G}_{2F}(u)$ below
\begin{align}
\mathbb{F}_{1F}(u)&=\frac{e^{\Phi_{F}(0,u)}\left[-b_{1}m-a_{2}\left(p_x{}-ip_y{}\right)\right]}{\sqrt{2}\,p_v{}}\,,\label{B6}\\\nonumber\\
\mathbb{G}_{2F}(u)&=\frac{e^{\Phi_{F}(0,u)}\left[-a_{2}m+b_{1}\left(p_x{}+ip_y{}\right)\right]}{\sqrt{2}\,p_v{}}\,.\label{B7}
\end{align}

\begin{remark}\label{norm}
\noindent We shall consider only positive energy solutions.  In addition, we shall omit the overall normalization factors from all solutions, as they will not play a role in our analysis.
\end{remark}

\noindent We write the solution in region $F$ in accordance with the notation of Eqs. \eqref{s1} and \eqref{B1}.  The positive energy solution in region $F$ is the usual Dirac plane wave in the chiral representation,
\begin{align}
\psi_{F}&=\left(\begin{array}{c}F_{1F}\\
F_{2F}\\
G_{1F}\\
G_{2F}
\end{array}\right)=e^{-i(p_x{}x+p_y{}y+p_v{}v)}
\left(\begin{array}{c}\mathbb{F}_1{}(u)\\
\mathbb{F}_2{}(u)\\
\mathbb{G}_1{}(u)\\
\mathbb{G}_2{}(u)
\end{array}\right),\label{B11}\\\nonumber\\
&=e^{-i(p_x{}x+p_y{}y+p_v{}v)}\left(\begin{array}{c}\displaystyle{\frac{\left(-b_{1}m-a_{2}\left(p_x{}-ip_y{}\right)\right)}{\sqrt{2}\,p_v{}}}\\\\
a2\\\\
b1\\\\
\displaystyle{\frac{\left(-a_{2}m+b_{1}\left(p_x{}+ip_y{}\right)\right)}{\sqrt{2}\,p_v{}}}
\end{array}\right)
e^{-ip_u{}u}.\label{B12}
\end{align}

\subsection[The solution in region $W$]{The solution in region $W$}\label{W}

\noindent For spacetime points in $W$, we have that $f(u)=\cos{(ku)}, \,g(u)=\cosh{(ku)}$.  As before, we assume that
\begin{equation}
\label{W0}
ka\in\left(-\frac{\pi}{2},\frac{\pi}{2}\right),
\end{equation}

\noindent in order to avoid the singularity arising from the vanishing of $f(u)$.  Then, following the same steps as in Sec. \ref{F}, we obtain,
\begin{equation}
\label{W1}
\Phi_{W}(0,u)=-iA(u)+\ln{C(u)},
\end{equation}

\noindent where $A(u)$ and $C(u)$ are given by
\begin{align}
A(u)&=\frac{\left(m^{2}ku+p_{x}^{2}\tan{(ku)}+p_{y}^{2}\tanh{(ku)}\right)}{2kp_v{}}\,,\label{W6a}\\\nonumber\\
C(u)&=\frac{1}{\sqrt{f(u)\,g(u)}}\,,\label{W6}
\end{align}

\noindent with the appropriate $f(u)$ and $g(u)$ from Eqs. \eqref{sw1} and \eqref{sw2}.

\noindent Then,
\begin{align}
\mathbb{F}_{1W}(u)&=\,e^{-iA(u)}C(u)\frac{\left[-b_{1}m-a_{2}p_x{}\sec{(ku)}+ia_{2}p_y{}\sech{(ku)}\right]}{\sqrt{2}\,p_v{}}\,,\label{W2}\\\nonumber\\
\mathbb{F}_{2W}(u)&=a_{2}\,e^{-iA(u)}C(u)\,,\label{W3}\\\nonumber\\
\mathbb{G}_{1W}(u)&=b_{1}\,e^{-iA(u)}C(u)\,,\label{W4}\\\nonumber\\
\mathbb{G}_{2W}(u)&=\,e^{-iA(u)}C(u)\frac{\left[-a_{2}m+b_{1}p_x{}\sec{(ku)}+ib_{1}p_y{}\sech{(ku)}\right]}{\sqrt{2}\,p_v{}}\,,\label{W5}
\end{align}\\

\noindent Thus 
\begin{equation}
\label{W7}
\psi_{W}=\left(\begin{array}{c}F_{1W}\\
F_{2W}\\
G_{1W}\\
G_{2W}
\end{array}\right)=e^{-i(p_x{}x+p_y{}y+p_v{}v)}
\left(\begin{array}{c}\mathbb{F}_{1W}(u)\\
\mathbb{F}_{2W}(u)\\
\mathbb{G}_{1W}(u)\\
\mathbb{G}_{2W}(u)
\end{array}\right).
\end{equation}\\

\noindent   On the boundary $u=0$, we have
\begin{align}
\mathbb{F}_{1W}(0)&=a_{1}=\frac{\left[-b_{1}m-a_{2}\left(p_x{}-ip_y{}\right)\right]}{\sqrt{2}\,p_v{}}\,,\label{W8}\\\nonumber\\
\mathbb{F}_{2W}(0)&=a_{2}\,,\label{W9}\\\nonumber\\
\mathbb{G}_{1W}(0)&=b_{1}\,,\label{W10}\\\nonumber\\
\mathbb{G}_{2W}(0)&=b_{2}=\frac{\left[-a_{2}m+b_{1}\left(p_x{}+ip_y{}\right)\right]}{\sqrt{2}\,p_v{}}\,,\label{W11}
\end{align}

\noindent as expected (see Eqs. \eqref{B6}, \eqref{B7}).  Furthermore one can check in a straightforward manner that the solution on the boundary between $F$ and $W$,  $\partial(FW)$, is $C^{1}$.

\subsection[The solution in region $B$]{The solution in region $B$}\label{B}

\noindent   In region $B$ we have the singularity due to the vanishing of $f(u)$ at $u=u_{s}$, where
\begin{equation}
\label{F0}
u_{s}=a+\frac{\cot{(ka)}}{k}\,.
\end{equation}

\begin{remark}\label{remarksing}
\noindent The singularity at $u=u_{s}$ is in fact a quasiregular singularity \textup{\cite{GP09}, \cite{ES77}}.  This singularity is a manifestation of the fact that no complete spacelike hypersurface exists that is adequate for the global specification of Cauchy data.  We also remark that the singularity moves along with the wave in the positive z-direction as time increases, \textup{(see Figure \ref{GSWFig1}}\textup{)}.
\end{remark}

\noindent The singularity at $u=u_{s}$, is present in the tetrad field and ultimately, as we shall see below, in the solution of the Dirac equation.\\

\noindent For region $B$, we follow the same basic procedure as in Sections \ref{F} and \ref{W}.  We can write, for all $a\leq u<u_{s}$,
\begin{equation}
\label{F1}
\Phi_{B}(0,u)=\Phi_{W}(0,a)+\Phi_{B}(a,u),
\end{equation}

\noindent where $\Phi_{W}(0,a)$, is obtained from Eq. \eqref{W1}.  We have
\begin{align}
\Phi_{W}(0,a)&=-iA(a)+\ln{C(a)},
\label{F2a}\\\nonumber\\
\Phi_{B}(a,u)&=\frac{i(a-u)}{2p_v{}}\left(m^{2}+\frac{p_x{}^{2}\sec{(ka)}}{f(u)}+\frac{p_y{}^{2}\sech{(ka)}}{g(u)}\right)\nonumber\\
&-\ln{C(a)}-\ln{\left(\sqrt{f(u)g(u)}\right)}\label{F2b}.
\end{align}

\noindent Thus
\begin{align}
\Phi_{B}(0,u)&=\frac{i(a-u)}{2p_v{}}\left(m^{2}+\frac{p_x{}^{2}\sec{(ka)}}{f(u)}+\frac{p_y{}^{2}\sech{(ka)}}{g(u)}\right)\nonumber\\
&-iA(a)-\ln{\left(\sqrt{f(u)g(u)}\right)}\label{F2bc}\,.
\end{align}

\noindent In Eq. \eqref{F2b}, and below, the $f(u)$ and $g(u)$, are given in Eqs. \eqref{sw1} and \eqref{sw2}.  For convenience we let
\begin{equation}
\label{F2e}
B(u)=\frac{i(a-u)}{2p_v{}}\left(m^{2}+\frac{p_x{}^{2}\sec{(ka)}}{f(u)}+\frac{p_y{}^{2}\sech{(ka)}}{g(u)}\right)-iA(a)\,,
\end{equation}

\noindent with $A(a)$ given by Eq. \eqref{W6a} and $C(u)$ by Eq. \eqref{W6} with the appropriate $f(u)$ and $g(u)$.  Thus we have,
\begin{align}
\mathbb{F}_{1B}(u)&=C(u)\,e^{B(u)}\left(\frac{1}{\sqrt{2}\,p_v{}}\right)\left(-b_{1}m-\frac{a_{2}p_x{}}{f(u)}+\frac{ia_{2}p_y{}}{g(u)}\right)\,,\label{F3}\\\nonumber\\
\mathbb{F}_{2B}(u)&=a_{2}C(u)\,e^{B(u)}\,,\label{F4}\\\nonumber\\
\mathbb{G}_{1B}(u)&=b_{1}C(u)\,e^{B(u)}\,,\label{F5}\\\nonumber\\
\mathbb{G}_{2B}(u)&=C(u)\,e^{B(u)}\left(\frac{1}{\sqrt{2}\,p_v{}}\right)\left(-a_{2}m+\frac{b_{1}p_x{}}{f(u)}+\frac{ib_{1}p_y{}}{g(u)}\right)\,.\label{F6}
\end{align}

\noindent Therefore,
\begin{equation}
\label{F7}
\psi_{B}=\left(\begin{array}{c}F_{1B}\\
F_{2B}\\
G_{1B}\\
G_{2B}
\end{array}\right)=e^{-i(p_x{}x+p_y{}y+p_v{}v)}
\left(\begin{array}{c}\mathbb{F}_{1B}(u)\\
\mathbb{F}_{2B}(u)\\
\mathbb{G}_{1B}(u)\\
\mathbb{G}_{2B}(u)
\end{array}\right).
\end{equation}

\noindent It is easy to verify using Eqs. \eqref{W2}-\eqref{W5} that on the boundary $u=a$, we have
\begin{align}
\mathbb{F}_{1B}(a)&=\mathbb{F}_{1W}(a)\,,\label{F8}\\\nonumber\\
\mathbb{F}_{2B}(a)&=\mathbb{F}_{2W}(a)\,,\label{F9}\\\nonumber\\
\mathbb{G}_{1B}(a)&=\mathbb{G}_{1W}(a)\,,\label{F10}\\\nonumber\\
\mathbb{G}_{2B}(a)&=\mathbb{G}_{2W}(a)\,.\label{F11}
\end{align}

\noindent  In this case again one can check that the match on the $\partial(WB)$ is $C^{1}$.  We note that the spinor components have branch points at $u=u_{s}$ and in general, $\mathbb{F}_{1B}(u)=\kappa\,\mathbb{G}_{2B}(u)=O(f(u)^{-3/2}),$ as $u\rightarrow u_{s}$, where $\kappa$ is some constant (see Eqs. \eqref{W6}, \eqref{F3}, and \eqref{F6}).

\section[Removal of the coordinate singularity]{The Fermi Observer} \label{sing}

\noindent  Recall that region $B$ is flat.\footnote{ Since the metric is only $C^{1}$ on the boundary separating regions $B$ and $W$ in the global coordinate system $\{t,x,y,z\}$, the boundary of $B$ must be excluded here.}  Since the Riemann curvature tensor $R_{\alpha\beta\gamma\delta}$ is identically zero in the interior of region $B$ except at the singularity,  $R_{\alpha\beta\gamma\delta}$ can be extended as the zero tensor to all of the interior of $B$. Therefore, there are Minkowski coordinate charts that cover the (open) interior of $B$. By transforming to Minkowski coordinates in $B$, the singularity in the metric in region $B$  can thus be eliminated by a coordinate transformation.\\

\noindent In this section we construct a consistent Minkowski tetrad field for a Fermi observer (lab frame) whose spatially-fixed timelike path extends from region $F$ through the wave pulse $W$ to region $B$.  All other Minkowski tetrad fields in region $B$ are then related to this one via Lorentz transformations.   To proceed, we consider the timelike path,

\begin{equation}
\label{FT1}
\gamma(\tau)=(t(\tau),x(\tau),y(\tau),z(\tau))=(\tau,0,0,0),\;\;\;\tau \equiv T\geq 0.
\end{equation}
It is easily checked that the path $\gamma(\tau)$ is geodesic, and it is also easily verified that the following orthonormal tetrad is parallel (i.e., parallel transported) along $\gamma(\tau)$: 
\begin{align}
\bar{e}_0{}&=(1,0,0,0)\, =\partial_t\,,\label{FA4}\\
\bar{e}_1{}&=(0,\frac{1}{f(u)},0,0)\,=\frac{1}{f(u)}\partial_x\,,\label{FA5}\\
\bar{e}_2{}&=(0,0,\frac{1}{g(u)},0)\,=\frac{1}{g(u)}\partial_y\,,\label{FA6}\\
\bar{e}_3{}&=(0,0,0,1)\, =\partial_z\,,\label{FA7}
\end{align}
where $u$ is given by Eq. \eqref{m2}. Recalling that $f(u) = g(u) = 1$ in region $F$, we see that a representative Minkowski tetrad (at $\gamma(0)$) from region $F$ is parallel transported to region $B$ for all $\tau > \sqrt{2}a$.  We define our Fermi observer to be $\gamma(\tau)$, and we refer to the Fermi coordinates in this context as the `laboratory frame' (see Fig. \ref{GSWFig1}).  As we show in the remainder of this section, and in Appendix \ref{Fermi App}, the Fermi coordinates restricted to an open set in region $B$ are Minkowski coordinates, and it will also be shown that the coordinate transformation may be extended to the entire interior of region $B$.  Associated with this coordinate system is a canonical Minkowski tetrad field in the interior of region $B$.\\

\noindent In order to display the Fermi coordinate transformations, it is convenient first to introduce the following notation,
\begin{align}
f(u)&=p+qu\,,\label{F12}\\
g(u)&=r+su\,,\label{F13}
\end{align}

\noindent where,
\begin{align}
p&=\cos{(ka)}+ka\sin{(ka)}\,,\label{F14}\\
q&=-k\sin{(ka)}\,,\label{F15}\\
r&=\cosh{(ka)}-ka\sinh{(ka)}\,,\label{F16}\\
s&=k\sinh{(ka)}\,.\label{F17}
\end{align}

\noindent In what follows, we designate the Fermi coordinates restricted to region $B$ by $(T,X,Y,Z)$.  Details of the following calculation from $(t,x,y,z)$ to exact Fermi coordinates $(T,X,Y,Z)$ are given in Appendix \ref{Fermi App}.\begin{align}
x&=\frac{X}{\zeta(T,Z)}\,,\label{FT2}\\
y&=\frac{Y}{\xi(T,Z)}\,,\label{FT3}\\
z&=Z-\frac{qX^{2}}{2\sqrt{2}\,\zeta(T,Z)}-\frac{sY^{2}}{2\sqrt{2}\,\xi(T,Z)}\,,\label{FT4}\\
t&=T-\frac{qX^{2}}{2\sqrt{2}\,\zeta(T,Z)}-\frac{sY^{2}}{2\sqrt{2}\,\xi(T,Z)}\,,\label{FT5}
\end{align}
\noindent where in Eqs. \eqref{FT2}-\eqref{FT5},
\begin{align}
\zeta(T,Z)&=p+q\left(\frac{T-Z}{\sqrt{2}}\right),\label{FT6}\\\nonumber\\
\xi(T,Z)&=r+s\left(\frac{T-Z}{\sqrt{2}}\right).\label{FT7}
\end{align}

\noindent The inverse of the transformation of Eqs. \eqref{FT2}-\eqref{FT5}  is given by,
\begin{align}
X&=x\,f(t,z)\,,\label{FT7.1}\\\nonumber\\
Y&=y\,g(t,z)\,,\label{FT7.2}\\\nonumber\\
Z&=z+\frac{1}{4}\left(\sqrt{2}\left(pqx^{2}+rsy^{2}\right)+\left(q^{2}x^{2}+s^{2}y^{2}\right)(t-z)\right)\,,\label{FT7.3}\\\nonumber\\
T&=t+\frac{1}{4}\left(\sqrt{2}\left(pqx^{2}+rsy^{2}\right)+\left(q^{2}x^{2}+s^{2}y^{2}\right)(t-z)\right)\,,\label{FT7.4}
\end{align}

\noindent where
\begin{align}
f(t,z)&=p+q\left(\frac{t-z}{\sqrt{2}}\right)\,,\label{F7.5}\\\nonumber\\
g(t,z)&=r+s\left(\frac{t-z}{\sqrt{2}}\right)\,.\label{F7.6}
\end{align}

\noindent We note that since the inverse transformation given by Eqs. \eqref{FT7.1}-\eqref{FT7.4} consists of quadratic and cubic polynomials in $(t,x,y,z)$, they are also Taylor expansions for the Fermi coordinates.  It is readily verified that these equations also follow from the general Taylor expansion formula for Fermi coordinates given by \textup{Eq. (25)} in reference \textup{\cite{KC08}}, which verifies the correctness of our calculations.

\begin{remark} \label{tvsT} The two time coordinates $t$ and $T$ are equal along the Fermi observer's timelike path $(t,0,0,0)$, but differ away from the path according to Eq.\eqref{FT7.4}.  As a consequence, they generate distinct foliations of an open neighborhood of the lab frame path.  A general feature of Fermi coordinates is that the \textit{a priori} time coordinate and the Fermi time coordinate are not equal.  As a further illustration, it is shown in \textup{\cite{KR18}} and references therein that cosmological coordinate time decreases to zero \textup{(}the instant of the big bang\textup{)}, on all hypersurfaces of constant Fermi time in Robertson-Walker cosmologies.  
\end{remark}

\noindent The identity, $t-z=T-Z$, follows from either set of transformation formulas, so region $B$ may be characterized by $t-z=T-Z \geq\sqrt{2}\,a$.  The transformation formulas in \textup{Eqs.} \eqref{FT2}-\eqref{FT5}  are valid at all points in region $B$ except for the case $T-Z= \sqrt{2}\,u_{s}$, where the singularity $u_s$ is given by \textup{Eq.} \eqref{F0}.  However, the inverse transformation formulas of \textup{Eqs.} \eqref{FT7.1}-\eqref{FT7.4} are valid even for $t-z=\sqrt{2}\,u_{s}$, i.e., they are valid at the singular points in region $B$.  This is because those transformation formulas are polynomials.  Thus, the Fermi/Minkowski coordinates $(T,X,Y,Z)$ are valid in all points of the interior of region $B$, as discussed in the first paragraph of this section.\\

\noindent Using the transformation \eqref{FT2}-\eqref{FT5}, the metric
\begin{equation}
\label{FT8}
ds^{2}=dt^{2}-dz^{2}-f^{2}(u)dx^{2}-g^{2}(u)dy^{2}\,,
\end{equation}

\noindent with the $f(u)$ and $g(u)$ given in Eqs. \eqref{F12}-\eqref{F13} for region $B$, is transformed into the Minkowski metric,
\begin{equation}
\label{FT9}
ds^{2}=dT^{2}-dX^{2}-dY^{2}-dZ^{2}\,.
\end{equation}

\begin{remark}\label{NStimelike}
Our Fermi observer, or lab frame, is one of an infinite set of timelike geodesics of the form,
\begin{equation}
\label{tg}
x(\tau)=C_x,\;\;\;y(\tau)=C_y,\;\;\;z(\tau)=C_{z},\;\;\;t(\tau)=\tau,
\end{equation}
where the constants $C_x,C_y,C_z$ may be chosen arbitrarily.  This is the set of timelike geodesic paths that are stationary relative to any particle whose $3$-momentum is zero in the global coordinate system $(t,x,y,z)$.  Referring to Eqs. \eqref{B2}, \eqref{B3a}, \eqref{B3b}, we note that for a Dirac particle with zero $3$-momentum, we must choose $p_x=p_y=0$ and $p_v=1/\sqrt{2}$.  A calculation shows that the tetrad field given by Eqs. \eqref{FA4}-\eqref{FA7} is parallel along each of these geodesic paths.  Moreover, each such timelike geodesic passes through the singularity $u=u_{s}$ where it is well defined and smooth. Fermi coordinates for each such path are Minkowskian in the interior of region $B$ and defines a canonical Minkowski tetrad field.  We chose $C_x=C_y=C_z=0$ for $\gamma(\tau)$ in Eq. \eqref{FT1} only for convenience.
\end{remark}

\section[Removal of the coordinate singularity]{Spinor solution in the Fermi tetrad field} \label{sing2}

\noindent The goal of this section is to transform our spinor solution to the Dirac equation, $\psi_B$, in region $B$, to its representation, $\Psi$, in Fermi coordinates $(T,X,Y,Z)$ with the associated (canonical) Minkowski tetrad field on region $B$, Eqs. \eqref{FT27}-\eqref{FT30}.  To do this, we first need to find the local Lorentz transformation $\Lambda$ that represents the rotation between the tetrad axes for $\{t,x,y,z\}$ and the tetrad axes for $(T,X,Y,Z)$ at each spacetime point in region $B$ (but we carry this out by using 1-forms instead of vectors).\\

\noindent The local Lorentz transformation $\Lambda$ determines a transformation matrix $L$ that acts on spinors (for details see Appendix \ref{L}).  The two matrices are related by the consistency relations,
\begin{equation}
\label{FT34'}
L^{-1}\gamma^A{}L=\Lambda^A{}_B{}\gamma^B{}\,,
\end{equation}

\noindent and
\begin{equation}
\label{FT35'}
\gamma^0{}L^{\dagger}\gamma^0{}=L^{-1}\,,
\end{equation}
 where the $\gamma$ matrices are given in the chiral representation in Appendix \ref{Chiral}. We note that Eq. \eqref{FT35'} is explained and motivated in \cite{PT09}, (cf., Eq. (5.396) page 246). We then have the relationship $\Psi=L\cdot\psi_{B\text{Fermi}}$.\\ 
 
\noindent We begin with the tetrad 1-forms in the region $B$ for the metric \eqref{FT8}, 
\begin{align}
\bar{e}^0{}&=(1,0,0,0)\, = dt\,,\label{FT10}\\
\bar{e}^1{}&=(0,f(t,z),0,0)\, = f(t,z)dx\,,\label{FT11}\\
\bar{e}^2{}&=(0,0,g(t,z),0)\, = g(t,z)dy\,,\label{FT12}\\
\bar{e}^3{}&=(0,0,0,1)\, = dz\,.\label{FT13}
\end{align}

\noindent The 1-forms Eqs. \eqref{FT10}-\eqref{FT13} are transformed, under the coordinate transformation \eqref{FT2}-\eqref{FT5}, into the the tetrad 1-forms, $h^A{}=(h^A{}_T{},h^A{}_X{},h^A{}_Y{},h^A{}_Z{})$, below
\begin{align}
h^0{}&=\left(1+\frac{1}{2}\left(\mathbb{A}^{2}+\mathbb{B}^{2}\right),\,-\mathbb{A},\,-\mathbb{B},\,-\frac{1}{2}\left(\mathbb{A}^{2}+\mathbb{B}^{2}\right)\right)\,,\label{FT14}\\\nonumber\\
h^1{}&=\left(-\mathbb{A},\,1,\,0,\,\mathbb{A}\right)\,,\label{FT15}\\\nonumber\\
h^2{}&=\left(-\mathbb{B},\,0,\,1,\,\mathbb{B}\right)\,,\label{FT16}\\\nonumber\\
h^3{}&=\left(\frac{1}{2}\left(\mathbb{A}^{2}+\mathbb{B}^{2}\right),\,-\mathbb{A},\,-\mathbb{B},\,1-\frac{1}{2}\left(\mathbb{A}^{2}+\mathbb{B}^{2}\right)\right),\label{FT17}
\end{align}

\noindent where the components are the coefficients in respective order of $dT, dX, dY, dZ$, and
\begin{align}
\mathbb{A}&=\frac{qX}{\sqrt{2}\,\zeta(T,Z)}\,,\label{FT18}\\\nonumber\\
\mathbb{B}&=\frac{sY}{\sqrt{2}\,\xi(T,Z)}\,.\label{FT19}
\end{align}

\noindent Spinors transform like scalars (0-forms) under coordinate transformations, \cite{PT09}, thus we can easily obtain the transformed spinor, $\psi_{BFermi}(T,X,Y,Z)$, in Fermi coordinates.  We let
\begin{equation}
\label{FT20}
\mathbb{D}(T,Z)=\frac{1}{\sqrt{\zeta(T,Z)\,\xi(T,Z)}}\,,
\end{equation}

\noindent and
\begin{align}
&H=\left(\frac{1}{2\,pv}\right)\left(m^{2}+\frac{p_x{}^{2}}{(p+qa)\zeta(T,Z)}+\frac{p_y{}^{2}}{(r+sa)\xi(T,Z)}\right)\left(\frac{T-Z}{\sqrt{2}}-a\right)\nonumber\\\nonumber\\
&+\frac{p_x{}X}{\zeta(T,Z)}+\frac{p_y{}Y}{\xi(T,Z)}+p_v{}\left(\frac{T+Z}{\sqrt{2}}-\frac{qX^{2}}{2\zeta(T,Z)}-\frac{sY^{2}}{2\xi(T,Z)}\right)+A\,,\label{FT21}
\end{align}\\

\noindent where, of course, $H=H(T,X,Y,Z)$, and
\begin{equation}
\label{FT21a}
A=\left(\frac{1}{2kp_v{}}\right)\left(m^{2}ka-\frac{p_{x}^{2}\,q}{k(p+qa)}+\frac{p_{y}^{2}\,s}{k(r+sa)}\right).
\end{equation}

\noindent The $A$ above, in Eq. \eqref{FT21a}, is the $A(a)$ of Eq. \eqref{W6a} re-expressed in terms of the constants $p,q,r,s$, (see Eqs. \eqref{F14}-\eqref{F17}).  Then, the wavefunction $\psi_{BFermi}$, that is, Eq. \eqref{F7}, but in Fermi $(T,X,Y,Z)$ coordinates, is given by
\begin{equation}
\label{FT22}
\psi_{BFermi}(T,X,Y,Z)=\left(\begin{array}{c}F_{1}\\
F_{2}\\
G_{1}\\
G_{2}
\end{array}\right),
\end{equation}

\noindent where
\begin{align}
F_{1}&=\frac{\mathbb{D}(T,Z)}{\sqrt{2}\,p_v{}}\,e^{-iH(T,X,Y,Z)}\left(-b_{1}m-\frac{a_{2}p_x{}}{\zeta(T,Z)}+\frac{ia_{2}p_y{}}{\xi(T,Z)}\right)\,,\label{FT23}\\\nonumber\\
F_{2}&=a_{2}\mathbb{D}(T,Z)\,e^{-iH(T,X,Y,Z)}\,,\label{FT24}\\\nonumber\\
G_{1}&=b_{1}\mathbb{D}(T,Z)\,e^{-iH(T,X,Y,Z)}\,,\label{FT25}\\\nonumber\\
G_{2}&=\frac{\mathbb{D}(T,Z)}{\sqrt{2}\,p_v{}}\,e^{-iH(T,X,Y,Z)}\left(-a_{2}m+\frac{b_{1}p_x{}}{\zeta(T,Z)}+\frac{ib_{1}p_y{}}{\xi(T,Z)}\right)\,,\label{FT26}
\end{align}

\noindent We may now transform the tetrad 1-forms Eqs. \eqref{FT14}-\eqref{FT17}, into the canonical 1-form fields,
\begin{align}
\hat{e}^0{}&=(1,0,0,0)\,= dT\,,\label{FT27}\\
\hat{e}^1{}&=(0,1,0,0)\,= dX\,,\label{FT28}\\
\hat{e}^2{}&=(0,0,1,0)\,= dY\,,\label{FT29}\\
\hat{e}^3{}&=(0,0,0,1)\,= dZ,\,\,\label{FT30}
\end{align}

\noindent using the local Lorentz transformation below (see e.g., ref. \cite{PT09}, Eq. (5.234).  Note that our notation differs from that of Parker and Toms)
\begin{equation}
\label{FT31}
\hat{e}^A{}_\alpha{}=\Lambda^A{}_B{}\,h^B{}_\alpha{}\,,
\end{equation}

\noindent where $\Lambda$ is a local Lorentz transformation which in our case is

\begin{equation}
\label{FT32}
\Lambda=\left( \begin{array}{cccc} 
1+\frac{1}{2}\left(\mathbb{A}^{2}+\mathbb{B}^{2}\right) & \mathbb{A} & \mathbb{B} & -\frac{1}{2}\left(\mathbb{A}^{2}+\mathbb{B}^{2}\right)\\
\rule{0in}{5ex}
	\mathbb{A} & 1 & 
        0 & 
        -\mathbb{A}\\
\rule{0in}{5ex}
	\mathbb{B} & 
        0 & 
        1 & 
        -\mathbb{B}\\
\rule{0in}{5ex}
	\frac{1}{2}\left(\mathbb{A}^{2}+\mathbb{B}^{2}\right) & 
        \mathbb{A} & 
        \mathbb{B} & 
        1-\frac{1}{2}\left(\mathbb{A}^{2}+\mathbb{B}^{2}\right)\\
\end{array} \right)\,.
\end{equation}\\

\noindent We let $\Psi$, be the spinor in region $B$, in Fermi coordinates but with the Fermi (canonical) tetrad, Eqs. \eqref{FT27}-\eqref{FT30}. Thus $\Psi$ is defined in a region of Minkowski spacetime (region $B$).  The corresponding matrix $L$, relates the spinors, $\Psi$ and $\psi_{BFermi}$, that is, $\Psi=L\cdot\psi_{BFermi}$.  The calculation of $L$ is given in Appendix \ref{L} and the result is,

\begin{equation}
\label{FT33}
L=\left( \begin{array}{cccc} 
1 & \mathbb{A}-i\mathbb{B} & 0 & 0\\
\rule{0in}{5ex}
	0 & 1 & 
        0 & 
        0\\
\rule{0in}{5ex}
	0 & 
        0 & 
        1 & 
        0\\
\rule{0in}{5ex}
	0 & 
        0 & 
        -\mathbb{A}-i\mathbb{B} & 
        1\\
\end{array} \right)\,.
\end{equation}\\

\noindent It is easy to check that $L$ and $\Lambda$ satisfy the consistency relations
\begin{equation}
\label{FT34}
L^{-1}\gamma^A{}L=\Lambda^A{}_B{}\gamma^B{}\,,
\end{equation}

\noindent and
\begin{equation}
\label{FT35}
\gamma^0{}L^{\dagger}\gamma^0{}=L^{-1}\,,
\end{equation}

\noindent in the chiral representation (see Appendix \ref{Chiral}).  We give, $\Psi$, below in a convenient shorthand notation using Eqs. \eqref{FT22} and \eqref{FT33},
\begin{equation}
\label{FT36}
\Psi=\left(\begin{array}{c}F_{1}+(\mathbb{A}-i\mathbb{B})F_{2}\\
F_{2}\\
G_{1}\\
G_{2}-(\mathbb{A}+i\mathbb{B})G_{1}
\end{array}\right).
\end{equation}

\noindent One may verify directly that both solutions, Eqs. \eqref{FT22} and \eqref{FT36}, satisfy the Dirac equations with their respective tetrads (e.g., using the Fock-Ivanenko coefficient approach).

\section[Spin polarizartion tests]{Spin polarization tests} \label{spin}

\noindent It is known that in a general semi-Riemannian spacetime we cannot define a conserved spin operator \cite{R01}.  Nevertheless, in the situation discussed in the present paper, our Dirac particle is in flat (Minkowski) spacetime both before (region $F$) and after the passage of the shock wave (region $B$).\footnote {We recall that at this point that the singularity manifests itself only through the square root branch point in the function $\mathbb{D}$ given by Eq. \eqref{FT20}.} It therefore follows that we can define spin operators \cite{O11}, $\Sigma^{A}$, corresponding to polarization in the positive $X,Y,Z,$ directions, for the positive energy solutions of spin $\frac{1}{2}$ particles, in Minkowski spacetime $(T,X,Y,Z)$, in the particle's rest frame.  These are, apart from a factor of $\frac{1}{2}$,
\begin{equation}
\label{S1}
\Sigma^{A}=\gamma^5{}\gamma^0{}\gamma^A{},\;\;\;\;\;A=1,2,3\equiv X,Y,Z.
\end{equation}

\noindent The $\Sigma^{A}$ of Eq. \eqref{S1} are correct provided our tetrad vectors are the canonical  ones for Minkowski spacetime, namely, Eqs. \eqref{FT27}-\eqref{FT30}.  However, we must restrict our analysis in region $B$ away from the singularity so that,
\begin{equation}
\label{S2}
\sqrt{2}\,a\leq t-z=T-Z\neq\sqrt{2}\,u_{s}.
\end{equation}

\noindent As mentioned above, the polarization of the wavefunction is determined by the choice of the constants $a_{2}$ and $b_{1}$.  The details on such matters can be found in references \cite{PS95} and \cite{S13}, keeping in mind that our chiral representation of the $\gamma$ matrices differs somewhat from theirs, (see Appendix \ref{Chiral}).

\subsection[Transformation Properties of the Spin Operator]{Transformation Properties of the Spin Operator}\label{TransProp}

\noindent In this subsection we consider Minkowski spacetime (or an open submanifold of Minkowski spacetime such as region $B$) with Minkowski coordinates $(T,X,Y,Z)$. We define spin operators, $\Sigma^{A}$ as in \cite{O11}, corresponding to polarization in the positive $X,Y,Z,$ directions, for the positive energy solutions of spin $\frac{1}{2}$ particles in the particle's rest frame. These are, apart from a factor of $\frac{1}{2}$, given by,
\begin{equation}
\label{S1}
\Sigma^{A}=\gamma^5{}\gamma^0{}\gamma^A{},\;\;\;\;\;A=1,2,3.
\end{equation}
We show below how the operator $\Sigma^{A}$ transforms under a (local) field of Lorentz transformations $\Lambda$ acting on the Minkowski tetrad field associated with the Minkowski coordinates $(T,X,Y,Z)$, and we find two equivalent expressions for the corresponding transformation of the operator $\Sigma^{A}$.\\

\noindent The $\Sigma^{A}$ of Eq. \eqref{S1} are spin operators in the particle's rest frame provided that our tetrad vectors are canonical for Minkowski spacetime.  For the purposes of this subsection only, we express the canonical tetrad field as:
\begin{align}
e_0{}&=(1,0,0,0)=\partial_T\,,\label{S2}\\
e_1{}&=(0,1,0,0)=\partial_X\,,\label{S3}\\
e_2{}&=(0,0,1,0)=\partial_Y\,,\label{S4}\\
e_3{}&=(0,0,0,1)=\partial_Z\,.\label{S5}
\end{align}
\noindent The above tetrads satisfy the orthonormality relations,
\begin{equation}
\label{S6}
e_A{}^\alpha{}e_B{}^\beta{}\eta_{\alpha\beta}=\eta_{AB}\,,
\end{equation}

\noindent where $\eta_{AB}$ is the Minkowski metric. If we now perform a local Lorentz transformation $\Lambda$ on the tetrad vectors \eqref{S2}-\eqref{S5}, we obtain a new set of tetrad vectors, $h_A{}^\alpha{}$, which are related to the ones above by the equation (written below in terms of the corresponding 1-forms)
\begin{equation}
\label{S7}
h^A{}_\alpha{}=\Lambda^A{}_B{}\,e^B{}_\alpha{}\,,
\end{equation}
\noindent where
\begin{equation}
\label{S8}
\Lambda^A{}_C{}\,\Lambda^B{}_D{}\,\eta_{AB}=\eta_{CD}\,,
\end{equation}

\noindent and, of course,
\begin{equation}
\label{S9}
h^A{}_\alpha{}\,h^B{}_\beta{}\,\eta_{AB}=\eta_{\alpha\beta}\,.
\end{equation}

\noindent The spacetime dependent $\gamma$ matrices \cite{PT09}, \cite{L76} are then given by,
\begin{equation}
\label{S10}
\bar{\gamma}^\alpha{}(x):=h_A{}^\alpha{}(x)\gamma^A{}\,.
\end{equation}

\noindent Using Eq. \eqref{S10} and $h^B{}_\alpha \,h_A{}^\alpha{}=\delta^B{}_A{}$, we obtain
\begin{equation}
\label{S22}
\gamma^B{}=h^B{}_\alpha{}\bar{\gamma}^\alpha{}\,.
\end{equation}

\noindent Then with the matrix $L$ related to $\Lambda$ {by Eq. \eqref{FT34}, we get, by rearranging terms,
\begin{align}
L\,\gamma^C{}L^{-1}&=\gamma^B{}\,\Lambda_B{}^C{}\,,\label{S23}\\
&=h^B{}_\alpha{}\,\Lambda_B{}^C{}\,\bar{\gamma}^\alpha{}.\label{S24}
\end{align}
\noindent But from Eq. \eqref{S7}, $h^B{}_\alpha{}=\Lambda^B{}_D{}\,e^D{}_\alpha{}$, so that,
\begin{align}
L\,\gamma^C{}L^{-1}&=\Lambda^B{}_D{}\,e^D{}_\alpha{}\,\Lambda_B{}^C{}\,\bar{\gamma}^\alpha{},\label{S26}\\
&=\delta_D{}^C{}\,e^D{}_\alpha{}\,\bar{\gamma}^\alpha{}.\label{S26}\\
&=e^C{}_\alpha{}\,\bar{\gamma}^\alpha{}:=\bar{\gamma}^C.\label{S27}
\end{align}
It now follows immediately that for each $A=X,Y,Z$,
\begin{equation}
\label{S11}
\overline{\Sigma}^{A}=\bar{\gamma}^5{}\bar{\gamma}^T{}\bar{\gamma}^A = (L\,\gamma^5{}L^{-1})(L\gamma^T L^{-1})(L\,\gamma^A{}L^{-1})=L\Sigma^{A}L^{-1}. 
\end{equation}
We summarize this results of this subsection in the form of the following theorem.

\begin{theorem}\label{theorem}
Let $\Psi$ be a spinor defined in a region of Minkowski space with coordinates $(T,X,Y,Z)$ with the canonical tetrad field given by Eqs. \eqref{S2}-\eqref{S5}.  Let $\Lambda$ be a local Lorentz transformation and let $\bar{\gamma}^\alpha{}(x)$ be defined by Eqs. \eqref{S10} and \eqref{S7}, and let $\overline{\Sigma}^{A}=:\bar{\gamma}^5{}\bar{\gamma}^T{}\bar{\gamma}^A $.  Suppose that $L$ and $\Lambda$ satisfy \eqref{FT34} and \eqref{FT35}.
Then for $A = X,Y$ or $Z$, 
\begin{equation}
\Sigma^{A}\Psi=\Psi \quad \text{if and only if}\quad \overline{\Sigma}^{A}L\Psi=L\Psi.
\end{equation}
\end{theorem}

\subsection[Spin tests in region $F$]{Spin tests in region $F$}\label{SF}

\noindent We begin this subsection by considering the situation in region $F$.  As previously mentioned we ignore overall normalization constants.  Of course the normalization will be the same in the different parts of the solution for each polarization case. Furthermore we shall also ignore the overall exponential factor. We denote by $u(p)$ the spinor part in Eq. \eqref{B12}, thus

\begin{equation}
\label{SF1}
u(p)=\left(\begin{array}{c}\displaystyle{\frac{\left(-b_{1}m-a_{2}\left(p_x{}-ip_y{}\right)\right)}{\sqrt{2}\,p_v{}}}\\\\
a2\\\\
b1\\\\
\displaystyle{\frac{\left(-a_{2}m+b_{1}\left(p_x{}+ip_y{}\right)\right)}{\sqrt{2}\,p_v{}}}
\end{array}\right)\,.
\end{equation}\\

\noindent For polarization in the positive $x$-direction we let
\begin{align}
a_{2}&=p_t{}+m-p_x{}-ip_y{}+p_z{}\,,\label{SF2}\\
b_{1}&=-p_t{}-m-p_x{}+ip_y{}-p_z{}\,,\label{SF3}
\end{align}

\noindent for polarization in the positive $y$-direction we let
\begin{align}
a_{2}&=i(p_t{}+m+ip_x{}-p_y{}+p_z{})\,,\label{SF4}\\
b_{1}&=-p_t{}-m-ip_x{}-p_y{}-p_z{}.\,\label{SF5}
\end{align}

\noindent Finally for polarization in the positive $z$-direction we let,
\begin{align}
a_{2}&=-\frac{p_x{}+ip_y{}}{p_t{}+m}\,,\label{SF6}\\
b_{1}&=-1-\frac{p_z{}}{p_t{}+m}\,.\label{SF7}
\end{align}

\noindent So in the ``lab frame,'' where, $p_x{}=p_y{}=p_z{}=0$, and $p_v{}=m/\sqrt{2}$, the spinor of Eq. \eqref{SF1} reduces to the three polarized polarized spinors below,\\
\begin{equation}
\label{SF8}
u_{x}(p)=\left(\begin{array}{c}2m\\\\
2m\\\\
-2m\\\\
-2m
\end{array}\right),\;\;\;\;
u_{y}(p)=\left(\begin{array}{c}2m\\\\
2im\\\\
-2m\\\\
-2im
\end{array}\right),\;\;\;\;
u_{z}(p)=\left(\begin{array}{c}1\\\\
0\\\\
-1\\\\
0
\end{array}\right).
\end{equation}\\

\noindent It is easy to check that
\begin{align}
&\Sigma^{1}\cdot u_{x}(p)-u_{x}(p)=0.\label{SF9}\\
&\Sigma^{2}\cdot u_{y}(p)-u_{y}(p)=0.\label{SF10}\\
&\Sigma^{3}\cdot u_{z}(p)-u_{z}(p)=0.\label{SF11}
\end{align}
Thus, the spinors in Eq. \eqref{SF8} are eigenvectors of the respective spin operators and are polarized in the positive $x,y$ and $z$ directions respectively.

\subsection[Spin tests in region $B$]{Spin tests in region $B$}\label{SB}

\noindent In this section we apply the tests of Eqs. \eqref{SF9}-\eqref{SF11} but use the $\Psi$ given by Eq. \eqref{FT36}.  In the ``lab frame'' we denote by $\Psi_{lab}$ the value of $\Psi$, when $p_x{}=p_y{}=p_z{}=0$, and $p_v{}=m/\sqrt{2}$.  The lab frame determines Fermi coordinates, which are also Minkowski coordinates, and the associated tetrad field (see Eqs. \eqref{S2}-\eqref{S5}).  The spinor solution $\Psi_{lab}$, whose momentum is specified and therefore has arbitrary location is given by,  

\begin{equation}
\label{SB1}
\Psi_{lab}=\frac{e ^{-\frac{im}{2}(2T-X\mathbb{A}-Y\mathbb{B})}}{\sqrt{\zeta(T,Z)\xi(T,Z)}}\left(\begin{array}{c}-b_{1}+a_{2}(\mathbb{A}-i\mathbb{B})\\
a_{2}\\
b_{1}\\
-a_{2}-b_{1}(\mathbb{A}+i\mathbb{B})
\end{array}\right),
\end{equation}

\noindent and where $\zeta,\xi,\mathbb{A},\mathbb{B}$ are given in Eqs. \eqref{FT6}, \eqref{FT7}, \eqref{FT18}, \eqref{FT19}, respectively.\\

\noindent To test the deviation from the initial $x$ polarization eigenstate we choose the $a_{2}$ and $b_{1}$ given by Eqs. \eqref{SF2} and \eqref{SF3}.  The notation $\Psi_{lab}^{(x)}$ means, initally polarized in the $x$ direction, and likewise for $\Psi_{lab}^{(y)}$ and $\Psi_{lab}^{(z)}$.  Then we find that
\begin{equation}
\label{SB2}
\Psi_{lab}^{(x)}=\frac{2m\,e ^{-\frac{im}{2}(2T-X\mathbb{A}-Y\mathbb{B})}}{\sqrt{\zeta(T,Z)\xi(T,Z)}}\left(\begin{array}{c}1+\mathbb{A}-i\mathbb{B}\\
1\\
-1\\
-1+\mathbb{A}+i\mathbb{B}
\end{array}\right),
\end{equation}

\noindent and that
\begin{equation}
\label{SB3}
\Sigma^{1}\cdot\Psi_{lab}^{(x)}-\Psi_{lab}^{(x)}\equiv \Delta\Psi_{lab}^{(x)}\neq 0,
\end{equation}

\noindent where
\begin{equation}
\label{SB4}
\Delta\Psi_{lab}^{(x)}=\frac{2m\,e ^{-\frac{im}{2}(2T-X\mathbb{A}-Y\mathbb{B})}}{\sqrt{\zeta(T,Z)\xi(T,Z)}}\left(\begin{array}{c}-\mathbb{A}+i\mathbb{B}\\
\;\;\,\mathbb{A}-i\mathbb{B}\\
\;\;\,\mathbb{A}+i\mathbb{B}\\
-\mathbb{A}-i\mathbb{B}
\end{array}\right).
\end{equation}

\noindent Likewise for initial polarization in the $y$ direction, we choose the $a_{2}$ and $b_{1}$ given by Eqs. \eqref{SF4} and \eqref{SF5} and obtain
\begin{equation}
\label{SB5}
\Psi_{lab}^{(y)}=\frac{2m\,e ^{-\frac{im}{2}(2T-X\mathbb{A}-Y\mathbb{B})}}{\sqrt{\zeta(T,Z)\xi(T,Z)}}\left(\begin{array}{c}1+i\mathbb{A}+\mathbb{B}\\
i\\
-1\\
-i+\mathbb{A}+i\mathbb{B}
\end{array}\right),
\end{equation}

\noindent and that
\begin{equation}
\label{SB6}
\Sigma^{2}\cdot\Psi_{lab}^{(y)}-\Psi_{lab}^{(y)}\equiv \Delta\Psi_{lab}^{(y)}\neq 0,
\end{equation}

\noindent where
\begin{equation}
\label{SB7}
\Delta\Psi_{lab}^{(y)}=\frac{2m\,e ^{-\frac{im}{2}(2T-X\mathbb{A}-Y\mathbb{B})}}{\sqrt{\zeta(T,Z)\xi(T,Z)}}\left(\begin{array}{c}-i\mathbb{A}-\mathbb{B}\\
\;\;-\mathbb{A}+i\mathbb{B}\\
-i\mathbb{A}+\mathbb{B}\\
\;\;-\mathbb{A}-i\mathbb{B}
\end{array}\right).
\end{equation}

\noindent Finally, to test the deviation from the $z$ polarization eigenstate we choose the $a_{2}$ and $b_{1}$ given by Eqs. \eqref{SF6} and \eqref{SF7}.  We find that
\begin{equation}
\label{SB8}
\Psi_{lab}^{(z)}=\frac{e ^{-\frac{im}{2}(2T-X\mathbb{A}-Y\mathbb{B})}}{\sqrt{\zeta(T,Z)\xi(T,Z)}}\left(\begin{array}{c}1\\
0\\
-1\\
\mathbb{A}+i\mathbb{B}
\end{array}\right),
\end{equation}

\noindent and that again
\begin{equation}
\label{SB9}
\Sigma^{3}\cdot\Psi_{lab}^{(z)}-\Psi_{lab}^{(z)}\equiv \Delta\Psi_{lab}^{(z)}\neq 0,
\end{equation}

\noindent where
\begin{equation}
\label{SB10}
\Delta\Psi_{lab}^{(z)}=-\frac{2\,e ^{-\frac{im}{2}(2T-X\mathbb{A}-Y\mathbb{B})}}{\sqrt{\zeta(T,Z)\xi(T,Z)}}\left(\begin{array}{c}0\\
0\\
0\\
\mathbb{A}+i\mathbb{B}
\end{array}\right)\neq 0.
\end{equation}

\begin{remark}\label{labspin} We note that the spin is unaltered only on the zero measure set of the Fermi observer's worldline, but the spatial position of the Dirac particle is completely unspecified so this is not a measurable phenomenon.  If a different laboratory frame described in Remark \ref{NStimelike} were used instead, the spinor along the original laboratory frame would be seen to be unpolarized. 
\end{remark}

\noindent We see that $\Sigma^{A}\Psi_{lab}^{(A)}\neq\Psi_{lab}^{A}$ for each coordinate direction $A$.  It now follows from Theorem \ref{theorem} that no Minkowski tetrad field exists in region $B$ with respect to which $\Psi_{lab}^{A}$, expressed in that tetrad field, is stationary and polarized in any coordinate direction.  To see this, assume to the contrary that there does exist such a Minkowski tetrad field.  Then that tetrad field must be related to the lab frame tetrad field through a Lorentz transformation $\Lambda$ with associated spinor representation $L$.  Using the notation of Theorem \ref{theorem}, it follows that
\begin{equation}\label{contradict}
\overline{\Sigma}^{A}L\Psi_{lab}^{A}=L\Psi_{lab}^{A},
\end{equation}
for some direction $A$ and at all spacetime points in region $B$. But this implies by Theorem \ref{theorem} that $\Sigma^{A}\Psi_{lab}^{A}=\Psi_{lab}^{A}$, which is a contradiction.\\

\noindent Thus, the effect of the gravitational wave pulse is to alter the spins so that the Dirac particle loses its initial polarization.  We note that we arrive at this same conclusion if any other timelike geodesic discussed in Remark \ref{NStimelike} is used to define a Minkowski tetrad field on the interior of region $B$ instead of the lab frame we used.  The calculations are essentially the same.

\section[Summary and concluding remarks]{Summary and concluding remarks} \label{conc}

\noindent In this paper, we considered a spacetime $(\mathcal{M}, \mathrm{g})$ for the gravitational wave pulse  (the `sandwich' wave of \cite{BPR59}), which  is a union of three regions, $F$ (front), $W$ (wave), and $B$ (back) as shown in Figure \ref{GSWFig1}.  We solved the Dirac equation for a particle in this spacetime.  We showed that starting with polarized Dirac particles (eigenstates of the operators $\Sigma^{A}$) in region $F$, regardless of the initial polarization state, the Dirac spinor cannot be an eigenfunction of the axial stationary spin operators in any possible Minkowski tetrad field in region $B$, after the passage of the gravitational wave pulse.  Our choice for a lab frame geodesic was arbitrary, as described in Remark \ref{NStimelike}, but this conclusion that the spin is unpolarized holds with any such choice because of Theorem \ref{theorem} and the remarks surrounding Eq \eqref{contradict}.   Our results are broadly consistent with those of \cite{BGO17}, which analyzed the behavior of \textit{classical} spinning particles polarized in advance of the arrival of a weak gravitational wave, and the effects of the wave on spin.  In our case, all of our solutions are exact, including the solution to the Dirac equation, the Fermi coordinate transformations, transformations of the tetrad field and spinor, and there was no linearization of the metric. 

\begin{appendices}

\section[The transformation to Fermi coordinates in region $B$]{The transformation to Fermi coordinates in region $B$}\label{Fermi App}

\noindent In this section we give the transformation to the exact Fermi coordinates, $(T,X,Y,Z)$. For further details see references \cite{KC08,KC3,CM06}. We take our Fermi observer to be the timelike geodesic,
\begin{equation}
\label{FA1}
\gamma(\tau)=(t(\tau),x(\tau),y(\tau),z(\tau))=(\tau,0,0,0),\;\;\;\tau=T\geq 0.
\end{equation}

\noindent We recall the metric Eq. \eqref{m1}
\begin{equation}
\label{FA2}
ds^{2}=dt^{2}-dz^{2}-f^{2}(u)dx^{2}-g^{2}(u)dy^{2}\,,
\end{equation}

\noindent where
\begin{equation}
\label{FA3}
u=\frac{t-z}{\sqrt{2}}.
\end{equation}

\noindent Our first step is to write the geodesic equations using the Lagrangian for the metric expressed in the form of Eq. \eqref{m4},
\begin{equation}
\label{g2}
L=\dot{v}\dot{u}-\frac{f^{2}(u)}{2}\dot{x}^{2}-\frac{g^{2}(u)}{2}\dot{y}^{2},
\end{equation}

\noindent where the overdot signifies derivative with respect to an affine parameter $\sigma$, and
\begin{align}
f(\sigma)&=p+qu(\sigma),\label{g3}\\
g(\sigma)&=r+su(\sigma),\label{g4}
\end{align}

\noindent where $p, q, r, s,$ are constants.  Recall that these constants were appropriately chosen so that the sandwich wave metric is $C^{1}$.\\

\noindent Lagrange's equations are
\begin{align}
p_x{}&=\frac{\partial L}{\partial \dot{x}}=-f^{2}(u)\dot{x}\,,\label{g5}\\\nonumber\\
p_y{}&=\frac{\partial L}{\partial \dot{y}}=-g^{2}(u)\dot{y}\,,\label{g6}\\\nonumber\\
p_v{}&=\frac{\partial L}{\partial \dot{v}}=\dot{u}\,,\label{g7}
\end{align}

\noindent where $p_x{}, p_y{}, p_v{},$ are constants (conserved canonical momenta).  Integrating Eq. \eqref{g7}, we obtain,
\begin{equation}
\label{g8}
u(\sigma)=p_v{}\sigma+u_{0}.
\end{equation}

\noindent From Eqs. \eqref{g5}, \eqref{g6} we have
\begin{align}
x(\sigma)&=-\int\frac{p_x{}}{f^{2}(u)}\,d\sigma=\frac{p_x{}}{p_v{}q\,(p+qu(\sigma))}+x_{0},,\label{g9}\\\nonumber\\
y(\sigma)&=-\int\frac{p_y{}}{f^{2}(u)}\,d\sigma=\frac{p_y{}}{p_v{}s\,(r+su(\sigma))}+y_{0}.\label{g10}
\end{align}

\noindent The easiest way to obtain $v(\sigma)$ is to substitute the solutions, \eqref{g8}-\eqref{g10}
in Eq. \eqref{g2}, for the different kinds of geodesics, and integrate the resulting ODE below
\begin{equation}
\label{g11}
2\dot{v}\dot{u}-f^{2}(\sigma)\dot{x}^{2}-g^{2}(\sigma)\dot{y}^{2}=
\begin{cases}
-1, &\textup{spacelike},\\
0, &\textup{null},\\
+1,&\textup{timelike}\,,\\ 
\end{cases}
\end{equation}

\noindent where, we have abused the notation slightly, so that $f(\sigma)=p+qu(\sigma)$, and so on. The results for the \textit{spacelike} case follow:
\begin{align}
x(\sigma)&=\frac{p_x{}}{p_v{}\,q\,f(\sigma)}+x_0{},\label{g12}\\\nonumber\\
y(\sigma)&=\frac{p_y{}}{p_v{}\,s\,g(\sigma)}+y_0{},\label{g13}\\\nonumber\\
v(\sigma)&=\left(\frac{-1}{2p_{v}^{2}}\right)\left(u(\sigma)+\frac{p_{x}^{2}}{q\,f(\sigma)}+\frac{p_{y}^{2}}{s\,g(\sigma)}\right)+v_{0},\label{g14}\\\nonumber\\
u(\sigma)&=p_v{}\,\sigma+u_0{}\,\label{g15}
\end{align}

\noindent The geodesics in $(x, y, z, t)$ coordinates follow from the relations,
\begin{align}
z(\sigma)&=\frac{v(\sigma)-u(\sigma)}{\sqrt{2}},\label{sp8}\\\nonumber\\
t(\sigma)&=\frac{v(\sigma)+u(\sigma)}{\sqrt{2}},\label{sp9}
\end{align}

\subsection[Spacelike geodesics]{Spacelike geodesics}\label{SG}

\noindent To obtain the Fermi coordinate transformation equations, we first write the general expressions for the spacelike geodesics in region $B$.  From the preceding calculations, we find,
\begin{align}
x(\sigma)&=\frac{p_x{}}{p_v{}\,q\,f(\sigma)}+x_0{},\label{sp1}\\\nonumber\\
y(\sigma)&=\frac{p_y{}}{p_v{}\,s\,g(\sigma)}+y_0{},\label{sp2}\\\nonumber\\
z(\sigma)&=\left(\frac{1}{\sqrt{2}}\right)\left(v_{0}-u(\sigma)-\left(\frac{1}{2p_{v}^{2}}\right)\left(u(\sigma)+\frac{p_{x}^{2}}{q\,f(\sigma)}+\frac{p_{y}^{2}}{s\,g(\sigma)}\right)\right),\label{sp3}\\\nonumber\\
t(\sigma)&=\left(\frac{1}{\sqrt{2}}\right)\left(v_{0}+u(\sigma)-\left(\frac{1}{2p_{v}^{2}}\right)\left(u(\sigma)+\frac{p_{x}^{2}}{q\,f(\sigma)}+\frac{p_{y}^{2}}{s\,g(\sigma)}\right)\right),\label{sp4}
\end{align}

\noindent where $\sigma$ is the proper length, $x_{0},y_{0},v_{0},u_{0},p_x{},p_y{},p_v{}$, are constants, and the functions $f(\sigma),\,g(\sigma),\,u(\sigma)$, are
\begin{align}
u(\sigma)&=p_v{}\,\sigma+u_0{},\label{sp5}\\
f(\sigma)&=p+qu(\sigma),\label{sp6}\\
g(\sigma)&=r+su(\sigma).\label{sp7}
\end{align}

\subsection[Fermi coordinate conditions]{Fermi coordinate conditions}\label{FC}

\noindent Our Fermi observer is the timelike geodesic
\begin{equation}
\label{fermi1}
\gamma(\tau)=(\tau,0,0,0).
\end{equation}
\noindent Fermi coordinates in region $B$ are uniquely determined by the parallel transport along $\gamma(\tau)$ of the tetrad of vectors given by Eqs. \eqref{FA4}-\eqref{FA7} to region $B$.  This parallel transported tetrad frame constitutes the Fermi coordinate axes.\\ 

\noindent At any point $Q$ on the Fermi observer worldline we may define a unit spacelike vector
\begin{equation}
\label{fermi2}
\mathbb{V}=\left(\dot{t}(0),\dot{x}(0),\dot{y}(0),\dot{z}(0)\right).
\end{equation}

\noindent The vector $\mathbb{V}$ is the tangent vector to a spacelike geodesic emanating from $Q$.  The constants, $x_{0}, y_{0}, v_{0},u_{0}$, in Eqs. \eqref{sp1}-\eqref{sp4} are chosen so that
\begin{equation}
\label{fermi3}
\left(t(0),x(0),y(0),z(0)\right)=(\tau,0,0,0)=\gamma(\tau).
\end{equation}

\noindent We find that
\begin{equation}
\label{fermi4}
\sqrt{2}\,u_{0}=\tau\equiv T.
\end{equation}

\noindent We now choose the constants $p_x{}, p_y{},p_v{}$, so as to satisfy the conditions below, Eqs. \eqref{fermi5}-\eqref{fermi8},
\begin{align}
f(0)\dot{x}(0)&=\frac{X}{\lambda},\label{fermi5}\\\nonumber\\
g(0)\dot{y}(0)&=\frac{Y}{\lambda},\label{fermi6}\\\nonumber\\
\dot{z}(0)&=\frac{Z}{\lambda},\label{fermi7}\\\nonumber\\
\dot{t}(0)&=0,\label{fermi8}
\end{align}

\noindent where $\dot{x}=dx/d\sigma$, etc., and
\begin{equation}
\label{fermi9}
\lambda=\sqrt{X^{2}+Y^{2}+Z^{2}}\,.
\end{equation}

\noindent Conditions \eqref{fermi5} and \eqref{fermi6} are satisfied by letting
\begin{align}
p_x{}&=-\frac{(p+qu_{0})X}{\lambda},\label{fermi10}\\\nonumber\\
p_y{}&=-\frac{(r+su_{0})Y}{\lambda},\label{fermi11}
\end{align}

\noindent Conditions \eqref{fermi7} and \eqref{fermi8} become
\begin{align}
\frac{-2p_{v}^{2}-\mathbb{C}}{2\sqrt{2}\,p_v{}}&=\frac{Z}{\lambda},\label{fermi12}\\\nonumber\\
\frac{2p_{v}^{2}-\mathbb{C}}{2\sqrt{2}\,p_v{}}&=0,\label{fermi13}
\end{align}

\noindent where
\begin{equation}
\label{fermi14}
\mathbb{C}=\left(\frac{Z}{\lambda}\right)^{2}\,.
\end{equation}

\noindent Condition \eqref{fermi13} is satisfied with
\begin{equation}
\label{fermi15}
p_v{}=\pm \sqrt{\frac{\mathbb{C}}{2}}\,,
\end{equation}

\noindent however, it is evident from Eq. \eqref{fermi12} that $p_v{}>0$ gives the negative of the right-hand-side, so that the correct choice is
\begin{equation}
\label{fermi16}
p_v{}=-\sqrt{\frac{\mathbb{C}}{2}}\,.
\end{equation}

\noindent In $(t,x,y,z)$ coordinates the Lagrangian $L$ is given by
\begin{equation}
\label{fermi17}
2L=\dot{t}^{2}-\dot{z}^{2}-f^{2}(u)\dot{x}^{2}-g^{2}(u)\dot{y}^{2}=-1.
\end{equation}

\noindent So using Eq. \eqref{fermi8} we have, as expected,

\begin{equation}
\label{fermi18}
f^{2}(0)\dot{x}^{2}(0)+g^{2}(0)\dot{y}^{2}(0)+\dot{z}^{2}(0)=1.
\end{equation}

\noindent Finally using Eq. \eqref{fermi4}, we find that the transformation relating the coordinates $(t, x, y, z)$ to the Fermi coordinates $(T, X, Y, Z)$ is given by
\begin{align}
x&=x(\lambda)=\frac{X}{\zeta(T,Z)}\,,\label{fermi19}\\\nonumber\\
y&=y(\lambda)=\frac{Y}{\xi(T,Z)}\,,\label{fermi20}\\\nonumber\\
z&=z(\lambda)=Z-\frac{qX^{2}}{2\sqrt{2}\,\zeta(T,Z)}-\frac{sY^{2}}{2\sqrt{2}\,\xi(T,Z)}\,,\label{fermi21}\\\nonumber\\
t&=t(\lambda)=T-\frac{qX^{2}}{2\sqrt{2}\,\zeta(T,Z)}-\frac{sY^{2}}{2\sqrt{2}\,\xi(T,Z)}\,,\label{fermi22}
\end{align}

\noindent where in Eqs. \eqref{fermi19}-\eqref{fermi22},
\begin{align}
\zeta(T,Z)&=p+q\left(\frac{T-Z}{\sqrt{2}}\right),\label{fermi23}\\\nonumber\\
\xi(T,Z)&=r+s\left(\frac{T-Z}{\sqrt{2}}\right).\label{fermi24}
\end{align}

\noindent Eqs. \eqref{fermi19}-\eqref{fermi22}, are Eqs. \eqref{FT2}-\eqref{FT5} in Sec. \ref{sing}, where we also give the inverse transformation, Eqs. \eqref{FT7.1}-\eqref{FT7.4}.  Using the transformation \eqref{fermi19}-\eqref{fermi22}, the metric
\begin{equation}
\label{fermi25}
ds^{2}=dt^{2}-dz^{2}-f^{2}(u)dx^{2}-g^{2}(u)dy^{2}\,,
\end{equation}

\noindent is transformed into the Minkowski metric,
\begin{equation}
\label{fermi26}
ds^{2}=dT^{2}-dX^{2}-dY^{2}-dZ^{2}\,.
\end{equation}

\section[The matrix $L$]{The matrix $L$}\label{L}

\noindent We use the equation below given in ref. \cite{PT09}, p. 225, with appropriate notational changes,
\begin{equation}
\label{L1}
L(\Lambda (x))=\exp{\left(i\epsilon_{AB}\Sigma^{AB}\right)}\,,
\end{equation}

\noindent where $\Lambda$ is a local Lorentz transformation, $\epsilon_{AB}$ are the parameters characterizing the Lorentz transformation.  Also from \cite{PT09}, Eq. (5.284), p. 228, we have that
\begin{equation}
\label{L2}
\Sigma^{AB}=-\frac{i}{8}\left[\gamma^A{},\gamma^B{}\right]\,,
\end{equation}

\noindent thus
\begin{equation}
\label{L3}
L=\exp{\left(\frac{\epsilon_{AB}}{8}\left[\gamma^{A},\gamma^B{}\right]\right)}
\end{equation}

\noindent Our $\gamma$ matrices will be in the chiral representation.\\

\noindent It is straightforward to show that, $\det{(L)}=1$.  We use the relation that for a matrix $M$ we have that
\begin{equation}
\label{L4}
\det{\left(e^{M}\right)}=e^{\tr{(M)}}\,.
\end{equation}

\noindent Now we use Eq. \eqref{L3} in its most general form.  The exponent then is
\begin{equation}
\label{L5}
\frac{1}{2}\left(\epsilon_{01}\gamma^{0}\gamma^{1}+\epsilon_{02}\gamma^{0}\gamma^{2}+\epsilon_{03}\gamma^{0}\gamma^{3}+\epsilon_{12}\gamma^{1}\gamma^{2}+\epsilon_{13}\gamma^{1}\gamma^{3}+\epsilon_{23}\gamma^{2}\gamma^{3}\right).
\end{equation}

\noindent By evaluating each product of pairs of $\gamma$ matrices separately in the chiral representation, we find that they all vanish.  Thus the trace of expression \eqref{L5} vanishes.  This is a representation independent result.\\

\noindent For the details of the relation between $\Lambda$ and $L$, namely the calculation of the parameters $\epsilon_{AB}$, a clear presentation is given in Chapter 1 of Hitoshi Yamamoto's lecture notes \cite{Y12} where it may be seen that $\Lambda$ may be expressed as
\begin{equation}
\label{L6}
\Lambda=e^{\xi_{i}K_{i}+\theta_{i}L_{i}}\equiv e^{M}\,,\;\;\;\;\;i=1,2,3,
\end{equation}

\noindent where the matrices $K_{i}$ and $L_{i}$ are given in \cite{Y12}, furthermore we find that
\begin{equation}
\label{L7}
M=\left( \begin{array}{cccc} 
0 & \xi_{1} & 
        \xi_{2} & 
        \xi_{3}\\
\rule{0in}{5ex}
	\xi_{1} & 
        0 &  
        -\theta_{3} & 
        \theta_{2}\\
\rule{0in}{5ex}
	\xi_{2} & 
        \theta_{3} & 
        0 & 
        -\theta_{1}\\
\rule{0in}{5ex}
	\xi_{3} & 
         -\theta_{2} & 
        \theta_{1} & 
        0\\
\end{array} \right)=\left( \begin{array}{cccc} 
0 & \epsilon_{01} & 
        \epsilon_{02} & 
        \epsilon_{03}\\
\rule{0in}{5ex}
	\epsilon_{01} & 
        0 &  
        -\epsilon_{12} & 
       -\epsilon_{13}\\
\rule{0in}{5ex}
	\epsilon_{02} & 
        \epsilon_{12} & 
        0 & 
        -\epsilon_{23}\\
\rule{0in}{5ex}
	\epsilon_{03} & 
         \epsilon_{13} & 
        \epsilon_{23} & 
        0\\
\end{array} \right)\,,
\end{equation}

\noindent or
\begin{equation}
\label{L8}
\epsilon_{01}=\xi_{1},\;\epsilon_{02}=\xi_{2},\;\epsilon_{03}=\xi_{3},\;\epsilon_{12}=\theta_{3},\;\epsilon_{13}=-\theta_{2},\;\epsilon_{23}=\theta_{1}.
\end{equation}\\

\noindent Using Mathematica's MatrixLog[$\Lambda$] with our $\Lambda$, Eq. \eqref{FT21}, we obtain our expression for $M$.  Comparing coefficients with \eqref{L3}, we find the $\epsilon_{AB}$.  Finally Mathematica's MatrixExp[$M$] gives us $L$.

\section[Amplitude and polarization]{Amplitude and polarization}\label{Psi4}
\noindent The gravitational wave part in region $W$, Eqs. \eqref{sw1},\eqref{sw2}, is characterized by the, in general complex, curvature component
\begin{equation}
\label{psi1}
\Psi_{4}(u)=|\Psi_{4}|e^{i\theta}.
\end{equation}

\noindent In the Newman-Penrose formalism $\Psi_{4}$ is related to the Weyl tensor components by the relation
\begin{align}
\Psi_{4}&=-C_{\alpha\beta\gamma\delta}\,\overline{m}^\alpha{}n^\beta{}\,\overline{m}^\gamma{}n^\delta{}\,,\label{psi2}\\
&=-C_{\alpha\beta\gamma\delta}\,\lambda_4{}^\alpha{}\lambda_2{}^\beta{}\lambda_4{}^\gamma{}\lambda_2{}^\delta{}\,.\label{psi3}\\
&=-C_{4242}\,,\label{psi4}
\end{align}

\noindent We remark that since our spacetime is Ricci flat, the Weyl is equal to the Riemann tensor.\\

\noindent In Eq. \eqref{psi1} above, $|\Psi_{4}|$ is the amplitude of the wave and $\theta$ its polarization \cite{GP09}, p. 325.  If $\theta$ is constant then the wave is said to be linearly polarized.  In our case we have that
\begin{equation}
\label{psi5}
\Psi_{4}=k^{2}\,,
\end{equation}

\noindent thus our result for the amplitude of the wave agrees with Rindler's deduction, \cite{R86}, p. 170.
 
\section[The flat space solution]{The flat space solution}\label{Chiral}
\noindent The solution in region $F$ is the usual Dirac plane wave solution in the chiral representation.

\begin{remark}\label{matrices}
\noindent We choose the chiral representation $\gamma$ matrices
\begin{equation}
\label{A1}
\gamma^{0}=\left(
 \begin{matrix}
     0 &  -I_{2}\\
     \rule{0in}{2ex}
     -I_{2} & 0\\
 \end{matrix}
 \right),\;\;\;\;\gamma^{K}=\left(
 \begin{matrix}
     0 &  \sigma^{K}\\
     \rule{0in}{2ex}
     -\sigma^{K}& 0\\
 \end{matrix}
 \right),\;\;\;\; K=(1,2,3)\,,
\end{equation}

\noindent where the $\sigma^{K}$ are the standard Pauli matrices, and  $\gamma^{5}=i\,\gamma^{0}\gamma^{1}\gamma^{2}\gamma^{3}$.  We have
\begin{equation}
\label{A2}
\left(\gamma^{0}\right)^{2}=I\,,\hspace{0.5cm}\left(\gamma^{K}\right)^{2}=-I.
\end{equation}
\end{remark}

\noindent We write everything in $x,y,v,u$, and $p_x{},p_y{},p_v{},p_u{}$, so that $p_\mu{}x^\mu{}=p_x{}x+p_y{}y+p_v{}v+p_u{}u$.  Then, for example, the positive and negative energy solutions with spin along the $z$ axis, in the Feynman-St\"{u}ckelberg interpretation, i.e., $p_t{}>0$ everywhere, are given below apart from an overall normalization factor, $N(p)=\sqrt{(p_t{}+m)/4m}$, where, we follow the Lorentz invariant normalization of \cite{O11} and \cite{IZ80}, namely $\bar{u}(p,s)u(p,s^{\prime})=\delta_{s,s^{\prime}}$, etc.

\begin{align}
&\psi^{(+)(1)}=\left(\begin{array}{c}\frac{\sqrt{2}\left(m+\sqrt{2}\,p_u{}\right)}{\sqrt{2}\,m+p_v{}+p_u{}}\\\\
\frac{-\sqrt{2}\left(p_x{}+i\,p_y{}\right)}{\sqrt{2}\,m+p_v{}+p_u{}}\\\\
\! \! \!\frac{-\sqrt{2}\left(m+\sqrt{2}\,p_v{}\right)}{\sqrt{2}\,m+p_v{}+p_u{}}\\\\
\frac{-\sqrt{2}\left(p_x{}+i\,p_y{}\right)}{\sqrt{2}\,m+p_v{}+p_u{}}
\end{array}\right)
e^{-ip_\mu{}x^\mu{}},\;\;\;
\psi^{(+)(2)}=\left(\begin{array}{c}\frac{-\sqrt{2}\left(p_x{}-i\,p_y{}\right)}{\sqrt{2}\,m+p_v{}+p_u{}}\\\\
\frac{\sqrt{2}\left(m+\sqrt{2}\,p_v{}\right)}{\sqrt{2}\,m+p_v{}+p_u{}}\\\\
\! \! \! \;\frac{-\sqrt{2}\left(p_x{}-i\,p_y{}\right)}{\sqrt{2}\,m+p_v{}+p_u{}}\\\\
\frac{-\sqrt{2}\left(m+\sqrt{2}p_u{}\right)}{\sqrt{2}\,m+p_v{}+p_u{}}
\end{array}\right)
e^{-ip_\mu{}x^\mu{}},\nonumber\\\nonumber\\\label{A3}\\
&\psi^{(-)(1)}=\left(\begin{array}{c}\frac{\sqrt{2}\left(m+\sqrt{2}p_u{}\right)}{\sqrt{2}\,m+p_v{}+p_u{}}\\\\
\frac{-\sqrt{2}\left(p_x{}+i\,p_y{}\right)}{\sqrt{2}\,m+p_v{}+p_u{}}\\\\
\! \! \! \frac{\sqrt{2}\left(m+\sqrt{2}\,p_v{}\right)}{\sqrt{2}\,m+p_v{}+p_u{}}\\\\
\frac{\sqrt{2}\left(p_x{}+i\,p_y{}\right)}{\sqrt{2}\,m+p_v{}+p_u{}}
\end{array}\right)
e^{ip_\mu{}x^\mu{}},\;\;\;\;\;\;\;\;
\psi^{(-)(2)}=\left(\begin{array}{c}\frac{-\sqrt{2}\left(p_x{}-i\,p_y{}\right)}{\sqrt{2}\,m+p_v{}+p_u{}}\\\\
\frac{\sqrt{2}\left(m+\sqrt{2}\,p_v{}\right)}{\sqrt{2}\,m+p_v{}+p_u{}}\\\\
\! \! \!\frac{\sqrt{2}\left(p_x{}-i\,p_y{}\right)}{\sqrt{2}\,m+p_v{}+p_u{}}\\\\
\frac{\sqrt{2}\left(m+\sqrt{2}p_u{}\right)}{\sqrt{2}\,m+p_v{}+p_u{}}
\end{array}\right)e^{ip_\mu{}x^\mu{}}.\nonumber
\end{align}\\ 

\end{appendices}


\newpage

\begin{thebibliography}{99}

\bibitem{C76} S. Chandrasekhar, ``The solution of Dirac's equation in Kerr geometry,'' Proc. R. Soc. Lond. A. \textbf{349}, 571-575 (1976).

\bibitem{Carter} B. Carter and R. G. McLenaghan. ``Generalized total angular momentum operator for the Dirac equation in curved space-time'' Phys. Rev. D  \textbf{19}, 1093-1097 (1979).

\bibitem{P80} L. Parker, ``One-electron atom as a probe of spacetime curvature,'' Phys. Rev. D  \textbf{22}, 1922-1934 (1980).

\bibitem{S91} G. V. Shishkin, ``Some exact solutions of the Dirac equation in gravitational fields,'' Class. Quantum Grav.  \textbf{8}, 175-185 (1991).

\bibitem{Z96} A. Zecca, ``The Dirac equation in the Robertson-Walker space-time,'' J. Math. Phys.  \textbf{37}, 874-879 (1996).

\bibitem{FR09} F. Finster, M. Reintjes, ``The Dirac equation and the normalization of its solutions in a closed Friedmann-Robertson-Walker universe,'' Class. Quantum Grav.  \textbf{26}, 105021 (2009).  \href{http://arxiv.org/abs/0901.0602v4}{arXiv:0901.0602v4}

\bibitem{HP09} X. Huang, L. Parker, ``Hermiticity of the Dirac Hamiltonian in curved spacetime,'' Phys. Rev. D  \textbf{79}, 024020 (2009).

\bibitem{R17} C. R\"{o}ken, ``The Massive Dirac Equation in Kerr Geometry: Separability in Eddington-Finkelstein-Type Coordinates and Asymptotics,'' Gen . Relativ. Gravit. \textbf{49}, 1-23 (2017).  \href{http://arxiv.org/pdf/1506.08038.pdf}{arXiv:1506.08038v2}

\bibitem{PT09} L. Parker, D. Toms, \textit{Quantum Field Theory in Curved Spacetime} (Cambridge U. Press, Cambridge, 2009), p. 227.

\bibitem{D70I} W. G. Dixon, ``Dynamics of extended bodies in general relativity I,'' Proc. Roy. Soc. Lond. A. \textbf{314}, 499-527 (1970).

\bibitem{D70II} W. G. Dixon, ``Dynamics of extended bodies in general relativity II,'' Proc. Roy. Soc. Lond. A. \textbf{319}, 509-547 (1970).

\bibitem{D74III} W. G. Dixon, ``Dynamics of extended bodies in general relativity III,'' Phil. Trans. Roy. Soc. Lond. A. \textbf{277}, 59-119 (1974).

\bibitem{HN90} F. W. Hehl, W.-T. Ni, ``Inertial effects of a Dirac particle,'' Phys. Rev. D  \textbf{42}, 2045-2048 (1990).

\bibitem{S08} A. J. Silenko, ``Gravitational waves and spinning test particles,'' Acta Physica Polonica B Proceedings Supplement  \textbf{1}, 87-107 (2008).

\bibitem{MO13} B. Mashhoon, Y. N. Obukhov, ``Spin precession in inertial and gravitational fields,'' Phys. Rev. D  \textbf{88}, 064037 (2013).

\bibitem{OST13} Y. N. Obukhov, A. J. Silenko, O. V. Teryaev, ``Spin in an arbitrary gravitational field,'' Phys. Rev. D  \textbf{88}, 084014 (2013).

\bibitem{MS00} M. Mohseni, H. R. Sepangi, ``Gravitational waves and spinning test particles,'' Class. Quantum Grav.  \textbf{17}, 4615-4626 (2000).

\bibitem{M02} M. Mohseni, ``Spinning particles in gravitational wave spacetime,'' Phys. Lett. A \textbf{301}, 382-388 (2002).

\bibitem{BGO17} D. Bini, A. Geralico, A. Ortolan, ``Deviation and precession effects in the field of a weak gravitational wave,'' Phys. Rev. D  \textbf{95}, 104044 (2017).

\bibitem{R01} L. H. Ryder, ``Spin in special and general relativity'', in C. L\"{a}mmerzahl, C. W. F. Everitt, F. W. Hehl, (Eds.) \textit{Gyros, Clocks, Interferometers...\textup{:} Testing Relativistic Gravity in Space} (Springer-Verlag, Berlin, 2001).

\bibitem{BPR59} H. Bondi, F. A. E. Pirani, I. Robinson, ``Gravitational waves in general relativity III.  Exact plane waves,'' Proc. Roy. Soc. Lond. A. \textbf{251}, 519-533 (1959).

\bibitem{R86} W. Rindler, \textit{Essential Relativity: Special, General, and Cosmological} Revised 2nd Ed. (Oxford U. Press, Oxford, 1986).

\bibitem{BFI89} D. Bini, V. Ferrari, J. Ibanez, ``Finite-energy wave packets of gravitational radiation,'' Nuovo Cimento  \textbf{103 B}, 29-44 (1989).

\bibitem{BP89} H. Bondi, F. A. E. Pirani, ``Gravitational waves in general relativity XIII.  Caustic property of plane waves,'' Proc. R. Soc. Lond. A. \textbf{421}, 395-410 (1989).

\bibitem{R06} W. Rindler, \textit{Relativity: Special, General, and Cosmological} 2nd Ed. (Oxford U. Press, Oxford, 2006).

\bibitem{GP09} J. B. Griffiths, J. Podolsk\'{y}, \textit{Exact Space-Times in Einstein's General Relativity} (Cambridge U. Press, Cambridge, 2009).

\bibitem{GHP73} R. Geroch, A. Held, R. Penrose, ``A space-time calculus based on pairs of null directions,'' J. Math. Phys.  \textbf{14}, 874-881 (1973).

\bibitem{C92} S. Chandrasekhar, \textit{The Mathematical Theory of Black Holes} (Clarendon Press, Oxford, 1992).

\bibitem{ES77} G. F. R. Ellis, B. G. Schmidt, ``Singular space-times,'' Gen. Relativ. Gravit.  \textbf{8}, 915-953 (1977).

\bibitem{KC08} D. Klein, P. Collas, ``General transformation formulas for Fermi-Walker coordinates,'' Class. Quantum Grav.  \textbf{25}, 145019 (2008).

\bibitem{KR18} D. Klein, J. Reschke, ``Pre-big bang geometric extensions of inflationary cosmologies,'' Ann. Henri Poincar\'e  \textbf{19}, 565-606 (2018), DOI: 10.1007/s00023-017-0634-6 

\bibitem{O11} T. Ohlsson \textit{Relativistic Quantum Physics} (Cambridge U. Press, Cambridge, 2011), (Eq. (3.100), p. 57).

\bibitem{PS95} M. E. Peskin, D. V. Schroeder, \textit{An Introduction to Quantum Field Theory} (Perseus Books Pub., Reading, 1995).

\bibitem{S13} A. M. Steane, ``An introduction to spinors,''  1-23, 13 Dec. 2013.\\
\href{http://arxiv.org/abs/1312.3824}{arXiv:1312.3824v1[math-ph]}

\bibitem{KC3}  Klein, D., Collas, P.: Exact Fermi coordinates for a class of spacetimes, \textit{J. Math. Phys.} \textbf{51} 022501(10pp) (2010). (arXiv:\href{https://arxiv.org/abs/0912.2779}{math-ph/0912.2779})


\bibitem{CM06} C. Chicone, B. Mashhoon, ``Explicit Fermi coordinates and tidal dynamics in de Sitter and G\"{o}del spacetime,'' Phys. Rev. D \textbf{74}, 064019 (2006).

\bibitem{Y12} H. Yamamoto, ``Quantum Field Theory for Non-specialists,'' \href{http://epx.phys.tohoku.ac.jp/~yhitoshi/particleweb/particle.html}{Unpublished Lecture notes, Ch. 1-6} (2012).

\bibitem{IZ80} C. Itzykson, J-B. Zuber, \textit{Quantum Field Theory} (McGraw-Hill Inc., New York, 1980).

\bibitem{L76} E. A. Lord, \textit{Tensors, Relativity and Cosmology} (Tata McGraw-Hill Publishing Co. Ltd., New Delhi, 1976).


\end{thebibliography}
\end{document}